\input ./style/arxiv-general.cfg
\documentclass[aoas,MSNbibl,nameyear,dvips]{arximspdf}
\makeatletter
   \@ifpackageloaded{graphicx}{}{\usepackage{graphicx}}
\makeatother
\doi{10.1214/15-AOAS832}% Updated by VTEXPTS2LaTeX.exe, 17.06.2015 08:27
\volume{9}
\issue{3}
\pubyear{2015}
\firstpage{1194}
\lastpage{1225}
\docsubty{FLA}

\makeatletter
\newcommand{\lleft}{\left}
\newcommand{\rrvert}{\vert}
\newcommand{\rright}{\right}
\newcommand{\rrVert}{\Vert}
\newcommand{\llvert}{\vert}
\newcommand{\llVert}{\Vert}
\newcommand{\veczero}{\mathbf{0}}
\newcommand{\vecx}{\mathbf{x}}
\newcommand{\vecy}{\mathbf{y}}

\newcommand{\vecp}{\mathbf{p}}
\newcommand{\vecP}{\mathbf{P}}

\newcommand{\vecX}{\mathbf{X}}
\newcommand{\vecY}{\mathbf{Y}}

\newcommand{\vecC}{\mathbf{C}}
\newcommand{\vecB}{\mathbf{B}}
\newcommand{\vecL}{\mathbf{L}}

\newcommand{\vecSigma}{{\bolds{\Sigma}}}
\newcommand{\vect}{\mathbf{t}}
\newcommand{\vecmu}{\bolds{\mu}}
\newcommand{\vecpsi}{\bolds{\psi}}
\newcommand{\vecxi}{\bolds{\xi}}
\newcommand{\vecdelta}{{\bolds{\delta}}}
\newcommand{\vecomega}{{\bolds{\omega}}}
\newcommand{\vecnu}{\bolds{\nu}}
\newcommand{\vecs}{\mathbf{s}}

\newcommand{\vectheta}{\bolds{\theta}}
\newcommand{\vecTheta}{\bolds{\Theta}}
\newcommand{\vecvartheta}{\bolds{\vartheta}}
\newcommand{\vecalpha}{\bolds{\alpha}}
\newcommand{\vecbeta}{\bolds{\beta}}
\newcommand{\vecgamma}{\bolds{\gamma}}
\newcommand{\veclambda}{\bolds{\lambda}}
\newcommand{\vecGamma}{\bolds{\Gamma}}
\newcommand{\reals}{\mathbb{R}}

\newcommand{\expec}{\mathbb{E}}
\newcommand{\PP}{\mathbb{P}}

\makeatother

\begin{document}
\begin{frontmatter}

\title{A Bayesian spatiotemporal model for reconstructing climate from
multiple pollen~records}
\runtitle{A Bayesian spatio-temporal temperature reconstruction}

\begin{aug}
% Corresponding author: Lasse Holmstr\"{o}m - lasse.holmstrom@oulu.fi% Updated by VTEXPTS2LaTeX.exe, 18.06.2015 12:31
%Updated by VTEXPTS2LaTeX.exe, 17.06.2015 08:27
\author[A]{\fnms{Lasse}~\snm{Holmstr\"{o}m}\corref{}\thanksref{m1}\ead[label=e1]{lasse.holmstrom@oulu.fi}},
\author[A]{\fnms{Liisa}~\snm{Ilvonen}\thanksref{m1}\ead[label=e2]{liisa.ilvonen@oulu.fi}},
\author[B]{\fnms{Heikki}~\snm{Sepp\"{a}}\thanksref{m2,T1}\ead[label=e3]{heikki.seppa@helsinki.fi}}
\and
\author[C]{\fnms{Siim}~\snm{Veski}\thanksref{m3}\ead[label=e4]{siim.veski@ttu.ee}}
\runauthor{Holmstr\"{o}m, Ilvonen, Sepp\"{a} and Veski}
\affiliation{University of Oulu\thanksmark{m1}, University of
Helsinki\thanksmark{m2} and\\ Tallinn University of Technology\thanksmark{m3}}
%\dedicated{}
\address[A]{L. Holmstr\"{o}m\\
L. Ilvonen\\
Department of Mathematical Sciences\\
University of Oulu\\
P.O. Box 3000\\
FIN-90014\\
Finland\\
\printead{e1}\\
\phantom{E-mail:\ }\printead*{e2}}
\address[B]{H. Sepp\"{a}\\
Department of Geosciences\\
\quad and Geography\\
University of Helsinki\\
P.O.Box 64\\
FIN-00014\\
Finland\\
\printead{e3}}
\address[C]{S. Veski\\
Institute of Geology\\
Tallinn University of Technology\\
Ehitajate tee 5\\
19086 Tallinn\\
Estonia\\
\printead{e4}}
\end{aug}
\thankstext{T1}{Supported by the Academy of Finland project Ecological History and Long-Term Dynamics of the
Boreal Forest Ecosystem (EBOR) and the Nordic top-level research initiative
Cryosphere-Atmosphere Interactions in a Changing Arctic Climate
(CRAICC).}

% HISTORY:
%
\received{\smonth{5} \syear{2014}}% Updated by VTEXPTS2LaTeX.exe,
%17.06.2015 08:27
%
\revised{\smonth{5} \syear{2015}}% Updated by VTEXPTS2LaTeX.exe,
%17.06.2015 08:27

% ABSTRACT
\begin{abstract}
Holocene (the last 12{,}000 years) temperature variation, including the
transition out of the last Ice Age to a warmer climate, is
reconstructed at multiple locations in southern Finland, Sweden and
Estonia based on pollen fossil data from lake sediment cores. A novel
Bayesian statistical approach is proposed that allows the reconstructed
temperature histories to interact through shared environmental response
parameters and spatial dependence. The prior distribution for past
temperatures is partially based on numerical climate simulation.
%The main patterns of the reconstructions are the marked rise of mean
%annual temperature from the early Holocene to the Holocene thermal
%maximum in northern Europe, followed by a gradual cooling towards the
%present. A brief cold episode is indicated 8200 years ago at two sites
%with particularly high sample resolution.
The features in the reconstructions are consistent with the
quantitative climate reconstructions based on more commonly used
reconstruction techniques. The results suggest that
the novel spatio-temporal approach can provide quantitative reconstructions
that are smoother, less uncertain and generally more realistic than the
site-specific individual reconstructions.
\end{abstract}

% KEYWORDS
\end{frontmatter}

\section{Introduction}

Instrumental temperature records rarely cover more than the past
100--200 years.
%In Finland, the longest measured time series are about 150 years long.
On the other hand,
temperature proxy data, such as fossil pollen, tree rings or ice cores,
provide a continuous and long record of climatic changes where
instrumental data do not exist [\citet{Janetal2007,Masetal2013}].
%%Temperature reconstructions based on such proxy data inform us of the
%degree of past temperature variation and occurrence of major climatic
%events.
The present article proposes Bayesian statistical methodology for
pollen-based paleotemperature reconstruction at multiple locations that
takes into account spatial and temporal dependencies between the sites
and along the cores. The method is then applied to reconstruct
Holocene, that is, post Ice Age mean annual temperature variation at
four locations in southern Finland, Sweden and Estonia based on fossil
pollen data extracted from lake sediment cores.

The standard approach to temperature reconstruction from multiple proxy
records is
the so-called Composite Plus Scaling (CPS) method
that uses the modern instrumental record and a suitable regression
technique to combine into an average representation the temperature
histories originally constructed only on the basis of the individual
records [e.g.,  \citet{JonesEtAl2009} and the references therein; see
also \citet{NRC}]. We propose a novel method that effectively combines
the data from all the original individual proxy records, in our case
the pollen taxon abundances, and reconstructs their temperature
histories in a joint estimation process that allows the histories to
interact through shared environmental response parameters and spatial
dependence. Our approach therefore represents a deeper integration of
the information in the proxy records than the standard methodology.

The usefulness of pollen and other organisms as temperature proxies is
based on the fact that different organisms tend to have different
optimal temperatures, that is, temperatures in which they fare
particularly well. Therefore, the relative abundances of different
pollen types in a sediment core layer reflect the temperature at the
time when the sediment layer was formed.
Pollen data is widely used in quantitative climate reconstructions
because pollen is abundant and widely dispersed and because the
importance of climate for the distribution and abundance of plants is
well studied and documented [\citet{Wood1987,Dahl1998}].
For recent reviews on climate reconstruction methodology, see
\citet{JonesEtAl2009,BirksEtAl2010,JugBir2012} and, for
pollen-based methods viewed from a Bayesian perspective, see \citet{OhlWah2012}.
%\citet{BirksEtAl2010} divides the existing reconstruction methods into
%three basic categories: the indicator-species approach, the assemblage
%approach, and the multivariate calibration-function approach.
%For every approach they consider assumptions made, strengths and
%weaknesses.
%As one example \cite{BirksEtAl2010} talk about
%As a strength of the Bayesian approach, \citet{BirksEtAl2010} and
%\citet{Bir2012} consider the use of researcher's a priori information
%about the parameters and the possibility to evaluate how well multiple
%working hypotheses fit data, instead of rejecting a null hypothesis.
%Moveover, the Bayesian approach allows advanced handling of
%uncertainty in data and model. The drawbacks include the lack of
%automatic software and long reconstruction times for some models. It
%is however the authors' view that
%All in all \cite{BirksEtAl2010} see that Bayesian methods can provide
%a rich way to develope climate reconstructions.
%a successful implementation of the Bayesian approach
%will be a major contribution to quantitative palaeoclimatology in
%the future.

%There are lot of papers about past climate reconstructions with
%different reconstruction methods. Below are some papers in which the
%method is similar with our method.

The Bayesian BARCAST model discussed in \citet{TingleyHuybers2010A} and
\citet{TingleyHuybers2010B} aims to reconstruct a spatially and
temporally complete climate process from incomplete proxy and
instrumental time series. The space--time covariance is assumed
separable and exponential in space. The prior model describes the
evolution of the true surface temperatures as a multivariate
autoregressive process with spatially correlated innovations.
% and the data level describes the relationship between measurements
%and the true field values. % and the prior level defines prior
%distributions for all unknown parameters.
The authors test their model by reconstructing North American surface
temperatures using an instrumental surface temperature data set, after
corrupting a number of time series to mimic proxy observations. The
results are also compared with those obtained using the regularized
expectation--maximization algorithm (RegEM) and it is concluded that a
Bayesian algorithm produces more skillful reconstructions as measured
by the coefficient of efficiency and the length of the uncertainty
intervals. %\cite{TingleyHuybers2010B} proceed comparison considering
%differing assumptions made by BARCAST and RegEM and also the impacts
%of these differences to the analysis. They conclude that BARCAST
%resulst in reconstructions are more skillful than those produced by
%RegEM as measured by the coefficient of efficiency.

\citet{LiEtAl2010} use Bayesian hierarchical modeling to reconstruct
past Northern Hemisphere mean temperatures. Their model combines
information from proxies with different temporal resolution and
forcings which act as external drivers of large-scale temperature evolution.
%As proxies they use tree rings, pollen abundances and borehole
%temperatures. External forcings include solar irradiance, volcanic
%aerosols and greenhouse gas concentration.
However, no real proxy data are used and, instead, the proxy records
are simulated on the basis of numerical climate model outputs. Further,
the model does not include a spatial component. The results of the
paper emphasize the
importance of information that reflects climate on a variety of frequencies.
%\cite{LiEtAl2010} test different combinations of proxies with
%different models (e.g. with/without forcing covariates, with/without
%proxy noise). They conclude that addititon of forcing covariates
%reduce rmse especially when there is no pollen data. With pollen data
%forcings have only slight significance because pollen data contain the
%variability information of decedal to centennial scale.

\citet{BrynjarsdottirBerliner2011} reconstruct ground surface
temperature histories with uncertainty estimates for the past 400 years
from nine borehole temperature records using Bayesian hierarchical
modeling. Temperature histories and heat flow parameters for boreholes
in the same region share the same mean and variance. To find out
whether the sharing of information across groups of data has any
influence, they also fit single-site models to each of the nine
boreholes and conclude that combining the boreholes in two subregions
allows the ground surface temperature history parameters to borrow
strength across boreholes.

\citet{TingleyEtAl2012} present an overview of the challenges in
inferring with uncertainties a climate process through space and time.
The authors propose a unifying Bayesian modeling and notational
framework for the paleoclimate reconstruction problem. As one advantage
of hierarchical modeling they view the possibility of constructing and
testing each component independently of the others before they are
incorporated into the hierarchy. %\cite{TingleyEtAl2012} also describe
%generally Bayesian hierarchical modelling for infering past climate.
%%As special cases of their model they mention for example BARCAST
%model of \cite{TingleyHuybers2010A} and a model of \cite{LiEtAl2010}.
%\cite{TingleyEtAl2012} point out that there are software package like
%WinBUGS (\cite{LunnEtAl2000}) which enables to fit a broad range of
%Bayesian models but usually for the hierarchical models the MCMC
%algorithms have to be constructed and implemented case by case.
%Moreover space-time models can be computationally burdensome.
%Therefore they ask for more collaboration between the climate science
%and statistics communities in order to develope more realistic models
%to climate reconstruction.

The method proposed in this article is directly related to the work of
\citet{VaskoEtAl2000,ToivonenEtAl2001} and \citet{KorholaEtAl2002}, who were the first to use
detailed Bayesian modeling for paleoclimate reconstruction from
assemblage data. \citet{ToivonenEtAl2001} introduced a Bayesian
response model called Bum based on a unimodal model for an organism's
response to temperature. \citet{VaskoEtAl2000} then further developed
the Bum model and introduced a Bayesian hierarchical multinomial
regression model that takes into account dependency between species.
This model was called Bummer and it was further analyzed and modified
in \citet{ErastoHolmstrom2006} and \citet{SalonenEtAl2012}.

The starting point of our approach is the Bummer model that we extend
in several important ways. As opposed to Bummer, our model handles
multiple proxy records and also takes into account their spatial
correlations. The cores can have different chronologies and the
reconstruction is performed on a common chronology obtained as their
union. Finally, instead of the simple i.i.d. normal model used in
Bummer, the temporal part of the temperature field prior is defined by
a multivariate Gaussian smoothing prior with the smoothing parameter
hyperprior elicited using numerical climate model simulation.
The shortcomings of the simple i.i.d. model Bummer model were
demonstrated in \citet{ErastoHolmstrom2006}.
%Their model considers just one core but we introduce an extented
%spatio-temporal multicore model. In our model the prior for past
%temperatures has a separable space-time covariance which is
%exponential in space and first-order smoothing in time. '

\citet{HaslettEtAl2006} also used hierarchical Bayesian modeling to
reconstruct the prehistoric climate at Glendalough in Ireland from
fossil pollen data. A single core is used for reconstruction and, as in
\citet{ErastoHolmstrom2006}, a temporally smoothing temperature prior
is used to reflect the fact that climate change can be assumed to
exhibit a degree of smoothness. The use of European-wide
pollen-vegetation-climate relationships led at some time intervals to
multimodality in the posterior distribution of the reconstructed
environmental variables.
%In the present article we limited the training set to Scandinavian and
%Baltic State environments and did not observe multimodality in the
%reconstructed mean annual temperature.
To avoid the multimodality typical to
continental training sets with multiple strong climatic gradients, we
limited the training set to Scandinavian and Baltic State environments where
a simple south--north temperature gradient is dominant.
%This temporally smoothing prior is defined via a random walk with
%increments based on $t_8$ distribution.

%J{\chr"E4}tet{\chr"E4}{\chr"E4}n pois artikkeli Hierarchical model
%building, fitting, and checking: A behind-the-scenes look at a
%Bayesian analysis of arsenic exposure pathways (Craigmile, Peter F.
%and Calder, Cathernine A. and Li, Hongfei and Paul, Rajib and Cressie,
%Noel).
Finally, \citet{OhlWah2012} interpret the model of
\citet{HaslettEtAl2006} as a Bayesian version of the so-called Modern
Analog Technique (MAT) and their general framework has similarities
with tour approach, too. The authors also discuss the special
challenges in pollen-based environmental reconstructions as well as the
appropriateness of the unimodal response model. Some of the model
elements used in \citet{Pac2009} are also similar to our proposal.

The rest of the paper is structured as follows. The model and its
various components are described in Section~\ref{model}. The data used
in the example reconstructions and the results obtained are presented
in Section~\ref{example}, and Section~\ref{discussion} summarizes our
main conclusions. An online supplement [\citet{Holetal14suppa}] includes
an analysis of the Gaussian taxon response model used, reference
reconstructions from Greenland ice cores and Scandinavian records,
additional reconstructions based on our pollen data, the core
chronologies, and charts of the sediment core pollen abundances for the
most important taxa used in temperature reconstructions. All data used
in this work are available in the online supplement
\citet{Holetal14suppb} and the Matlab code used in reconstructions is in the
online supplement \citet{Holetal14suppc}.

\section{The model}
\label{model}

\subsection{The Bayesian method}

Bayesian inference is based on Bayes' theorem, which in its simplest
form can be written as
\begin{equation}
\label{eq1} p(\vecTheta|\mathrm{data}) =\frac{p(\vecTheta)p(\mathrm{data}|\vecTheta
)}{p(\mathrm{data})} \propto p(
\vecTheta)p(\mathrm{data}|\vecTheta).
\end{equation}
Here ``$\mathrm{data}$'' consists of the available observations and in our
case includes training lake and sediment core pollen abundances as well
as modern temperatures at the training lakes. The model parameters as
well as the past unknown temperatures are included in $\vecTheta$. The density
$p(\mathrm{data}|\vecTheta)$ is the likelihood of the data, the prior
distribution $p(\vecTheta)$ describes our prior beliefs about the model
parameters, and $p(\vecTheta|\mathrm{data})$ is the posterior
distribution of $\vecTheta$. Using the posterior distribution, the
investigator can in principle answer any question about the
probabilities of the unknown quantities of interest. Additional levels
of hierarchy can be added to the model by assuming that the prior of
$\vecTheta$ depends on another parameter $\vecpsi$ which in turn has
its own prior $p(\vecpsi)$, etc. For more information on Bayesian
modeling, see, for example,  \citet{BanerjeeEtAl} and \citet{Gelman}.

\subsection{Notation}
\label{notation}

In the following, the symbols for ``modern'' (training) and sediment
fossil quantities have the superscript $m$ and $f$, respectively. We
assume $n$ training lakes with known modern temperatures and $C$ cores
with $l$ pollen taxa counted from the training lakes and $l_c$ taxa
counted from core $c = 1,\ldots,C$. All core taxa are present also in
the training lakes. For core $c$, the number of depths sampled is
$n_c$, indexed according to increasing sediment age. The term ``site''
refers either to a training lake or to a depth in a core. Therefore,
there are
$n + n_1 + \cdots+n_C$ sites altogether.

\subsubsection*{Training lakes}
\[
\begin{tabular}{@{}lp{258pt}@{}}
{$\vecx^m = [x^m_1,\ldots,x^m_n]^T$} &   modern training
temperatures (30-year annual means);\\
${\vecy}^m_i = [y^m_{i1},\ldots,y^m_{il}]^T$ &   scaled modern pollen
taxon abundances
$(y^m_{i\cdot}\equiv\sum_{j=1}^{l}y^m_{ij}=100)$ at the training
lake $i$, $i = 1,\ldots,n$; \\
${\vecY}^m = [{\vecy}^m_1,\ldots,{\vecy}^m_n]$ &  $l\times n$ matrix of
modern taxon abundances.
\end{tabular}
\]

\subsubsection*{Cores}
\[
\begin{tabular}{@{}lp{248pt}@{}}
$\vecx^f_c = [x^f_{c1},\ldots,x^f_{cn_c}]^T$ &  unknown\vspace*{2pt}
past temperatures for core $c$,
$c = 1,\ldots,C$; \\
$\vecX^f = \{\vecx^f_1,\ldots,\vecx^f_C\}$ &  set of\vspace*{2pt} all past
temperatures;  \\
$\vecy^f_{ci} = [y^f_{ci1},\ldots,y^f_{cil_c}]^T$ &  scaled pollen
taxon abundances
($y^f_{ci\cdot}\equiv\sum_{j=1}^{l_c}y^f_{cij}=100$)
for core $c$ at site (depth) $i$,
$i=1,\ldots,n_c$, $c = 1,\ldots,C$;\\
$\vecY^f_c = [\vecy^f_{c1},\ldots,\vecy^f_{cn_c}]$&   $l_c \times n_c$
matrix of taxon abundances for core $c$,
$c = 1,\ldots,C$; \\
$\vecY^f = \{\vecY^f_1,\ldots,\vecY^f_C\}$ &  set of all core taxon
abundances.
\end{tabular}
\]

\subsubsection*{Reconstruction times: Chronologies}

The past temperature $x^f_{ci}$ for core $c$ at depth $i$ corresponds
to a time $t_{ci}$ determined using radiocarbon or other dating
technique. The sequence $t_{c1} > \cdots> t_{cn_c}$ is referred to as
the chronology of core $c$. We will reconstruct the past temperature on
a time grid defined by the union of all such chronologies,
\[
\vect=\{t_1,\ldots,t_N\} = \bigcup
_{c=1}^C \bigcup_{i=1}^{n_c}
\{ t_{ci}\},
\]
where $t_1 > \cdots> t_N$.
Note that one may have $N < n_1+\cdots+n_c$, because different core
chronologies may include identical dates.
The same grid is used for each core, which means that for a given core,
pollen abundance data will not be available for all time points.
However, intra- and inter-core temperature correlations will help
estimate the corresponding\vspace*{1pt} past temperatures in a reasonable manner.
We use the notation
$\widetilde{\vecx}^f_c = [\widetilde{x}^f_{c1},\ldots,\widetilde
{x}^f_{cN}]^T$ for the past\vspace*{1pt} temperatures at core $c$ on this union
chronology and $\widetilde{\vecX}^f = [(\widetilde{\vecx}^f_1)^T,\ldots
,(\widetilde{\vecx}^f_C)^T]^T$ for the $\mathit{NC}$ dimensional vector
that contains the past temperatures on the union chronology grid for
all cores. Thus, $\vecx^f_c$ is a subset of $\widetilde{\vecx}^f_c$ and
$\vecX^f$ is a subset of $\widetilde{\vecX}^f$.

\subsection{A Bayesian multinomial Gaussian response model for multiple cores}
\label{multibummer}

Our starting point is the Bummer model introduced in \citet{VaskoEtAl2000}. We will first generalize it to multiple cores and then
propose a further extension that takes into account spatial and
temporal correlations among the cores (Sections~\ref{spat-tempmodel}
and \ref{spatest}).
% For previous applications of the Bummer model and its modifications,
%see \cite{KorholaEtAl2002}, \cite{ErastoHolmstrom2006} and
%\cite{SalonenEtAl2012}.

Our aim is to find the posterior density $p(\widetilde{\vecX}^f | \vecY
^f, \vecx^m, \vecY^m)$ of past temperatures $\widetilde{\vecX}^f$ given
data $\vecY^f, \vecx^m$ and $\vecY^m$. If
$\vectheta$ contains the parameters of the model, taking $\vecTheta= \{
\vecX^f,\vectheta\}$ and conditioning the probabilities on $\vecx^m$,
we get from (\ref{eq1}) that
\begin{eqnarray}
p\bigl(\widetilde{\vecX}^f | \vecY^f,
\vecx^m, \vecY^m\bigr) & = & \int p\bigl(\widetilde{
\vecX}^f,\vectheta| \vecY^f,\vecx^m,
\vecY^m\bigr) \,d\vectheta
\nonumber
\\[-8pt]
\label{posterior}
\\[-8pt]
\nonumber
& \propto & \int p\bigl(\widetilde{\vecX}^f,\vectheta|
\vecx^m\bigr) p\bigl(\vecY^f,\vecY ^m|
\vecx^m,\vecX^f,\vectheta\bigr) \, d\vectheta.
\end{eqnarray}
In practice, posterior inference on past temperatures is performed by
generating a sample from $p(\widetilde{\vecX}^f,\vectheta| \vecY
^f,\vecx^m,\vecY^m)$ and keeping the part
corresponding to $\widetilde{\vecX}^f$.

Sites are assumed to be conditionally independent given the
temperatures and model parameters and, therefore, the likelihood term
can be expanded as
\begin{eqnarray}
&& p\bigl(\vecY^f,\vecY^m|\vecx^m,
\vecX^f,\vectheta\bigr)
\nonumber
\\
\label{condindep} &&\qquad = p\bigl(\vecY^f|\vecX^f,
\vectheta\bigr) p\bigl(\vecY^m|\vecx^m,\vectheta\bigr) =
\prod_{c=1}^C p\bigl(\vecY^f_c|
\vecx^f_c,\vectheta\bigr) p\bigl(\vecY^m|
\vecx ^m,\vectheta\bigr)
\\
&& \qquad= \prod_{c=1}^C \prod
_{i=1}^{n_c}p\bigl(\vecy^f_{ci}|x^f_{ci},
\vectheta\bigr) \prod_{i=1}^np\bigl(
\vecy^m_i | x^m_i,\vectheta
\bigr),
\nonumber
\end{eqnarray}
where in the second equality the conditional independence of the cores
was assumed.
This is one of the assumptions made in the original Bummer model and
may well be
an oversimplification. We decided to adopt it in order to limit the
complexity of the model.

Each site is assumed to have its own set of taxon occurrence
probabilities that reflects the probability of observing the various
taxa at that site.
%There are $l$ taxa counted at each modern site and $l_c$ taxa counted
%for the core $c$ and $l_c\subseteq l$.
Let
$\vecp^m_1,\ldots,\vecp^m_n \in\reals^l$ be the taxon probabilities at
the modern sites and let
$\vecp^f_{c1},\ldots, \vecp^f_{cn_c} \in\reals^{l_c}$ be the
corresponding probabilities for core $c$.
Denote
\[
\vecP^m = \bigl[\vecp^m_1,\ldots,
\vecp^m_n\bigr], \qquad\vecP^f = \bigcup
_{c=1}^C\bigl\{\vecp^f_{c1},
\ldots, \vecp^f_{cn_c}\bigr\}.
\]
Following \citet{VaskoEtAl2000}, we use a Gaussian function to model
how pollen abundance responds to temperature. The unimodal shape of the
response is intended to reflect the fact that each pollen taxon host
plant is assumed to have an optimum temperature at which it fares
particularly well and that the favorability of the temperature declines
symmetrically around this optimum [cf. \citet{KorholaEtAl2002}]. For
taxon $j$ at modern site $i$ the response is characterized by
\begin{equation}
\label{modernlambda} \lambda^m_{ij} = \alpha_j \exp
\biggl[- \biggl(\frac{\beta_j - x^m_i}{\gamma
_j} \biggr)^2 \biggr], \qquad i=1,
\ldots,n, j = 1,\ldots,l,
\end{equation}
where $\alpha_j$ is a scaling factor, $\beta_j$ models the optimum
temperature for taxon $j$, and $\gamma_j$ is a tolerance parameter. %(cf.
%\ Figure~\ref{responsemodel}).
These parameters are assumed to be the same for both training data and
the cores. Therefore, the response for core $c$ is described by
\begin{equation}
\label{fossillambda} \lambda^f_{cij} = \alpha_{k(c,j)} \exp
\biggl[- \biggl(\frac{\beta_{k(c,j)}
- x^f_{ci}}{\gamma_{k(c,j)}} \biggr)^2 \biggr],
\end{equation}
where $i=1,\ldots,n_c$, $j = 1,\ldots,l_c$ and the indices
$k(c,1),\ldots,k(c,l_c)$ correspond to the taxa counted from core $c$.
Let $\vecalpha= [\alpha_1,\dots,\alpha_l]^T$, $\vecbeta= [\beta
_1,\ldots,\beta_l]^T$, $\vecgamma= [\gamma_1,\ldots,\gamma_l]^T$ and define
$
\vecalpha^f_c = [\alpha_{k(c,1)} ,\ldots,\alpha_{k(c,l_c)} ]^T$,
and similarly for $\vecbeta^f_c$ and $\vecgamma^f_c$. Thus, $\vecalpha
^f_c$, $\vecbeta^f_c$ and $\vecgamma^f_c$ are the subvectors of
$\vecalpha$, $\vecbeta$ and
$\vecgamma$ that correspond to those taxa that appear in core $c$.
All modern Gaussian response model parameters are now denoted by
$\vecvartheta^m = [\vecalpha,\vecbeta,\vecgamma]$ and the corresponding
parameters for core $c$ by
$\vecvartheta^f_c = [\vecalpha^f_c,\vecbeta^f_c,\vecgamma^f_c]$. The
parameter vector
$\vectheta$ above is then defined as $\vectheta= \{\vecP^m,\vecP
^f,\vecvartheta^m\}$.
%, where $\kappa_c$ is a core specific smoothing parameter to be
%discussed later.
%\begin{figure}
%\psfrag{a}{$\alpha_j$}
%\psfrag{b}{$\beta_j$}
%\psfrag{c}{$\beta_j-\gamma_j$}
%\psfrag{d}{$\beta_j+\gamma_j$}
%\centerline{
%\includegraphics[width=0.8\linewidth]{response_figure.eps}
%}
%\caption{The Gaussian taxon response model used (formulas (
%\ref{modernlambda}) and (\ref{fossillambda})).
%}
%\label{responsemodel}
%\end{figure}

Other environmental factors besides the temperature can affect pollen
taxon abundances and this is modeled by treating the taxon
probabilities as random variables that follow a Dirichlet distribution,
\begin{eqnarray}
\vecp_i^m | x^m_i,
\vecvartheta^m &\sim& \operatorname{Dirichlet}\bigl(\veclambda
^m_i\bigr),\qquad i = 1,\ldots,n,
\nonumber
\\[-8pt]
\label{Dir}
\\[-8pt]
\nonumber
\vecp^f_{ci} | x^f_{ci},
\vecvartheta^f_c &\sim& \operatorname{Dirichlet}\bigl(
\veclambda^f_{ci}\bigr), \qquad i = 1,\ldots,n_c,
c = 1,\ldots,C,
\nonumber
\end{eqnarray}
where $\veclambda_i^m=[\lambda^m_{i1},\lambda^m_{i2},\ldots,\lambda
^m_{il}]$ and
$\veclambda_{ci}^f=[\lambda^f_{ci1},\lambda^f_{ci2},\ldots,\lambda^f_{cil_c}]$.
Considering the full conditional distributions of the probability vectors
$\vecp^m_i$ and $\vecp^f_{ci}$, the components of $\veclambda_i^m$ and
$\veclambda_{ci}^f$ can be interpreted as ``pseudo counts'' that are
added to the actual observed taxon relative abundances (cf.
Appendix~\ref{conditionals}).
The observed scaled taxon abundances are assumed to follow multinomial
distributions with the probabilities $\vecp^m_i,\vecp^f_{ci}$,
\begin{eqnarray}
\vecy^m_i | x^m_i,\vectheta &
\sim& \operatorname{Mult}\bigl(y^m_{i\cdot},
\vecp^m_i\bigr), \qquad i = 1,\ldots,n,
\nonumber
\\[-8pt]
\label{multin}
\\[-8pt]
\nonumber
\vecy^f_{ci} | x^f_{ci},
\vectheta &\sim& \operatorname{Mult}\bigl(y^f_{ci\cdot}, \vecp
^f_{ci}\bigr), \qquad i = 1,\ldots,n_c, c = 1,
\ldots,C.
\end{eqnarray}
We note that because of the Dirichlet distribution used, the average
taxon probabilities (\ref{Dir}) are determined by the relative size
of the responses $\lambda^m_{ij}$ (or $\lambda^f_{cij}$). As a result,
the temperature dependent taxon probabilities in the model can assume
much more general shapes than just a simple Gaussian. This also means
that the interpretation of the parameters $\alpha_j$, $\beta_j$ and
$\gamma_j$ is not straightforward. This is discussed in more detail in
\citet{Holetal14suppa}.
%where the dot $\cdot$ denotes summation over the corresponding index.

The prior term in (\ref{posterior}) can be factored as
\begin{equation}
%\begin{eqnarray}
\label{prior} %\lefteqn{
p\bigl(\widetilde{\vecX}^f,
\vectheta|\vecx^m\bigr) %}
%& & \nonumber
%\\
%& & = p(\vecX^f,\vecP^m,\vecP^f,\vecvartheta^m, \kappa_c |\vecx^m)
%\nonumber\\
%& & = p(\vecP^m,\vecP^f | \vecX^f,\vecvartheta^m,\kappa_c,\vecx^m)p(
%\vecX^f,\kappa_c,\vecvartheta^m | \vecx^m) \nonumber\\
%& &
=
p\bigl(\vecP^m | \vecx^m, \vecvartheta^m
\bigr) p\bigl(\vecP^f | \vecX^f, \vecvartheta^m
\bigr) p\bigl(\widetilde{\vecX}^f,\vecvartheta^m|
\vecx^m\bigr), %\end{eqnarray}
\end{equation}
and further, by (\ref{Dir}),
%The first two factors on right the hand side of (\ref{prior}) can be
%written as (cf.\ \ref{Dir})
\begin{eqnarray}
\label{Dirmod} p\bigl(\vecP^m | \vecx^m,
\vecvartheta^m\bigr) &=& \prod_{i=1}^n
p\bigl(\vecp^m_i | x^m_i,
\vecvartheta^m\bigr) = \prod_{i=1}^n
\operatorname{Dirichlet}\bigl(\vecp^m_i |
\veclambda^m_i\bigr),
\\
\label{dub} p\bigl(\vecP^f | \vecX^f,
\vecvartheta^m\bigr) &=& \prod_{c=1}^C
\prod_{i=1}^{n_c} p\bigl(\vecp^f_{ci}
| x^f_{ci},\vecvartheta^f_c\bigr)
= \prod_{c=1}^C\prod
_{i=1}^{n_c} \operatorname{Dirichlet}\bigl(
\vecp^f_{ci} | \veclambda^f_{ci}
\bigr).
\end{eqnarray}
Here conditional independence of the probabilities, given the temperatures, was assumed.
Assuming that the taxon-specific parameters are mutually independent
\textit{a~priori},
the third factor on the right-hand side of (\ref{prior}) can be written as
\begin{equation}
%\begin{eqnarray}
\label{prior2} %\lefteqn{
p\bigl(\widetilde{\vecX}^f,
\vecvartheta^m|\vecx^m\bigr) %} & & \nonumber\\
%& & = p(\vecX^f,\kappa_c| \vecvartheta^m,\vecx^m) p( \vecvartheta^m|
%\vecx^m) \nonumber\\
%& & = p(\vecX^f,\kappa_c) p( \vecvartheta^m|\vecx^m) \nonumber\\
%& & = p(\vecX^f|\kappa_c) p(\kappa_c)p( \vecvartheta^m|\vecx^m)
%\nonumber\\
%& & = p(\vecX^f|\kappa_c)p(\kappa_c) p(\vecalpha) p(\vecbeta|\vecx^m)
%p(\vecgamma) \nonumber\\
%& &
=
p\bigl(\widetilde{\vecX}^f\bigr) \prod_{j=1}^l
p(\alpha_j) \prod_{j=1}^l p(
\beta _j) \prod_{j=1}^l p(
\gamma_j). %\end{eqnarray}
\end{equation}

The above model reduces to the original Bummer
if only a single core is considered ($C = 1$) and the priors in (\ref{prior2}) are specified appropriately. In particular, Bummer uses an
i.i.d. Gaussian prior for $\widetilde{\vecX}^f$ that does not model
temporal correlation between past temperatures. In the next section, a
spatio-temporal prior for $\widetilde{\vecX}^f$ is described.

\subsection{A spatio-temporal model for past temperatures}
\label{spat-tempmodel}

We now define the prior distributions on the right-hand side of (\ref
{prior2}). The priors of the scaling factor $\alpha_j$ and the
tolerance parameter $\gamma_j$ are specified analogously to
\citet{VaskoEtAl2000} and \citet{KorholaEtAl2002},
\[
\alpha_j \sim\operatorname{Unif}(0.1,60),\qquad
\gamma_j \sim \operatorname{Gamma}(9,1/3), \qquad j=1,\ldots,l.
\]
%
%Parameter $\alpha_j$ is regarded as persentages but in practice
%scaling of each taxon is considered only in relation to other taxa (
%\cite{KorholaEtAl2002}).

For the prior of the optimum taxon temperature $\beta_j$ of taxon $j$,
our approach is different from the original Bummer specification which
used a Gaussian prior centered on the modern temperature of the single
core lake used in temperature reconstruction. With several cores
involved and all of them located at one end of a large training set
temperature gradient (cf. Section~\ref{data}), it makes more sense to
work in the spirit of empirical Bayes analysis and define reasonable
priors with the help of information gleaned from the training data.
Thus, following the weighted-averaging partial least squares (WA-PLS)
modeling idea of \citet{terBraakJuggins1993},\vspace*{1pt} we first estimate the
optimal temperature by
$\hat{\beta_j}=(\sum_{i=1}^{n}y^m_{ij})^{-1}\sum_{i=1}^n y^m_{ij}x_i^m$,
where $x_i^m$ and $y^m_{ij}$ are the modern temperature and the
abundance of taxon $j$ for training lake $i$, respectively.
%, and $y^m_{\cdot j}$ is the total abundance of taxon $j$ in the whole
%training set.
%for training lake $i$, $x_i^m$ is the temperature and $y^m_{ij}$ is
%the abundance of taxon $j$, and $y^m_{j\cdot}$ is the total abundance
%of taxon $j$ in the training set.
The $\hat{\beta_j}$'s thus estimated vary between $-2.7^{\circ}$C and
6.2$^{\circ}$C.
A vague prior for $\beta_j$ is then defined as
\[
\beta_j \sim\mathrm{N}\bigl(\hat{\beta_j},(1.5
\sqrt{3})^2\bigr), \qquad j=1,\ldots,l.
\]
For more discussion on the choice of this particular prior, see \citet{SalonenEtAl2012}.

It remains to describe the prior distribution of the vector $\widetilde
{\vecX}^f$
% = [(\widetilde{\vecx}^f_1)^T,\ldots,(\widetilde{\vecx}^f_C)^T]^T$
that consists of the unknown past temperatures
$\widetilde{\vecx}^f_c = [\widetilde{x}^f_{c1},\ldots,\widetilde
{x}^f_{cN}]^T$ for all cores, defined
on the union chronology time grid $t_1 > \cdots> t_N$. The prior is a
multivariate Gaussian with
a separable covariance matrix obtained as the Kronecker product of
spatial and temporal covariances,
\begin{equation}
\label{Kronecker} \vecSigma= \vecC_S \otimes\vecC_T \in
\reals^{\mathit{CN}\times \mathit{CN}}.
\end{equation}
To estimate the $C \times C$ spatial covariance matrix $\vecC_S$,
two different approaches were tried. In the first, non-Bayesian
approach, we applied Estimated Generalized Least Squares to fit a
continuously indexed isotropic covariance function $C_S(\vecs,\vecs')$
to the training temperature residuals obtained after subtracting a
linear trend and then defined $\vecC_S = [C_S(\vecs_c,\vecs_{c'})]$,
where $\vecs_c,\vecs_{c'}$ are the core locations [e.g., \citet{Cressie}]. In the Bayesian approach, an isotropic spatial covariance
of temperatures was included in the hierarchical model as an additional
parameter with its own prior distribution (Section~\ref{spatest}). The
two methods lead to rather similar reconstructions and in the following
we will report only results obtained with the latter approach.

While the training data can be expected to inform us of the
correlations between temperatures at different locations,
we do not have any such direct knowledge of the \textit{past} temperatures
that could be used to specify the temporal covariance $\vecC_T$. We
therefore believe that one should not use a temporal prior that makes
too restrictive assumptions about the actual past temperature values
and instead use a prior that basically only describes their internal
variability or ``roughness''
[cf. \citet{ErastoHolmstrom2006}].

The matrix $\vecSigma$ in (\ref{Kronecker}) is a block matrix with
blocks $C_S(\vecs_c,\vecs_{c'})\vecC_T$ so that the temporal covariance
for each core is $C_S(\vecs_c,\vecs_{c})\vecC_T$ and,
because of stationarity, $C_S(\vecs_c,\vecs_{c})$ actually does not
depend on $c$.
To define
$\vecC_T$, we assume that it defines the dependence structure of a process
\begin{equation}
\label{depmodel} \widetilde{x}^f_{c(i+1)} =
\widetilde{x}^f_{ci} + \frac{1}{\sqrt{\kappa
}}(t_{i+1} -
t_i)\varepsilon_i,
\end{equation}
where the $\varepsilon_i$'s are independent standard Gaussian variables and
$\kappa> 0$.
Thus, if $\widetilde{\vecx}_c^f=[\widetilde{x}_{c1}^f,(\widetilde{\vecx
}_{c*}^f)^T]^T$, so that
$\widetilde{\vecx}_{c*}^f=[\widetilde{x}_{c2}^f,\ldots,\widetilde
{x}_{cN}^f]^T$, we have for a fixed $\widetilde{x}^f_{1}$ that
\begin{equation}
\label{conditionalprior} p\bigl(\widetilde{\vecx}_{c*}^f|
\widetilde{x}_{c1}^f,\kappa\bigr) \propto\kappa
^{({N-1})/{2}} \exp \Biggl[-\frac{\kappa}{2} \sum_{i=2}^{N}
\biggl(\frac
{\widetilde{x}_{ci}^f-\widetilde{x}_{c(i-1)}^f}{t_{i}-t_{i-1}} \biggr)^2 \Biggr].
\end{equation}
Assuming that $\widetilde{x}_{c1}^f \sim\mathrm{N}(\mu_c,1)$, where $\mu
_c$ is the modern temperature at core lake $c$, we have after some
matrix algebra (cf.
Appendix~\ref{algebra}) that
\begin{eqnarray}
%\label{properprior}
p\bigl(\widetilde{\vecx}_c^f|\kappa\bigr)
&=& p\bigl(\widetilde{\vecx}_{c*}^f|\widetilde
{x}_{c1}^f,\kappa\bigr)p\bigl(\widetilde{x}_{c1}^f
\bigr)
\nonumber
\\[-8pt]
\label{properprior}
\\[-8pt]
\nonumber
&\propto & \kappa^{({N-1})/{2}}\exp \bigl[ -\tfrac{1}{2}\bigl(
\widetilde{\vecx }_c^f - \vecmu_c
\bigr)^T\vecGamma^{-1}\bigl(\widetilde{\vecx}_c^f-
\vecmu_c\bigr) \bigr],
\end{eqnarray}
where
\[
\vecmu_c=[\mu_c,\ldots,\mu_c]^T
, \qquad \vecGamma %\begin{pmatrix}
%\kappa_cB_{21}^TB_{22}^{-1}B_{21}+\frac{1}{\sigma_c^2} &
%\kappa_cB_{21}^T \\
%\kappa_cB_{21} & \kappa_cB_{22}
%\end{pmatrix}^{-1}
=
\pmatrix{ \vecB_{21}^T\vecB_{22}^{-1}
\vecB_{21}\kappa+1 & \kappa\vecB_{21}^T
\vspace*{3pt}
\cr
\kappa\vecB_{21} & \kappa\vecB_{22}}^{-1}
\]
and where the matrices $\vecB_{21}$ and $\vecB_{22}$ are defined in
Appendix~\ref{algebra}.
Thus, for core $c$,
$\widetilde{\vecx}_c^f|\kappa\sim\mathrm{N}(\vecmu_c,\vecGamma)$ and
$\vecC_T = \vecC_T(\kappa) =\vecGamma$ in (\ref{Kronecker}).

\begin{figure}

\includegraphics{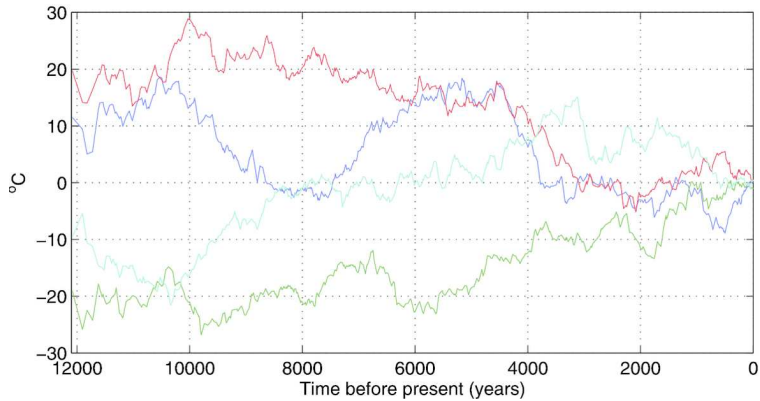}

\caption{Four realizations of past temperatures generated from their
prior distribution when the modern temperature is set at $\mu_c = 0$.}
\label{realizations}
\end{figure}

We see from (\ref{depmodel}) or (\ref{conditionalprior}) that $\kappa$
is a smoothing parameter that controls the roughness of past
temperature variation. A prior distribution will be specified for~$\kappa$ in Section~\ref{EstimationOfKappa}.
%This added level of hierarchy also means that the prior $p(\widetilde{
%\vecX}^f)$ in (\ref{prior2}) must be
%replaced by $p(\widetilde{\vecX}^f | \kappa)p(\kappa)$.
By definition,
\begin{equation}
\label{finalprior} \widetilde{\vecX}^f|\vecSigma\sim\mathrm{N}(\vecmu,
\vecSigma),
\end{equation}
with
\[
\vecmu=[\vecmu_1,\ldots,\vecmu_C]^T
\]
and $\vecmu_c=[\mu_c,\ldots,\mu_c]^T\in\reals^{N}$. To get an idea of
the nature of this prior,
Figure~\ref{realizations} shows sample paths from the marginal
distribution of $\mathrm{N}(\veczero,\vecSigma)$ that correspond to a
single core when $\kappa=306$, a point estimate suggested for
spatio-temporal reconstruction by the method of Section~\ref{EstimationOfKappa}.
As one can see, going back in time, the variance for past temperatures
grows rapidly,
making the prior very vague.
We also note that the prior favors rather slowly varying temperature
time series, limiting physically unreasonable fluctuations. This is
considered reasonable because the Holocene climate has been relatively
stationary.

%f

\subsection{The spatial covariance model}
\label{spatest}

Let $x^m(\vecs)$, $\vecs\in D$, denote the field of modern
temperatures in the region $D$ where the training lakes are located. We
assume that
\[
x^m(\vecs) = \mu(\vecs) + \delta(\vecs),
\]
where $\mu(\vecs)$ is a trend and $\delta(\vecs)$ is zero-mean and
isotropic. The covariance
function of $x^m(\vecs)$ is then
\[
C_S\bigl(\vecs,\vecs'\bigr) = \operatorname{Cov}
\bigl(x^m(\vecs),x^m\bigl(\vecs'\bigr)\bigr)
= \operatorname{Cov}\bigl(\delta (\vecs),\delta\bigl(\vecs'\bigr)
\bigr).
\]
%
%so that it is enough to estimate the covariance of the residual $
%\delta(\vecs)$.
In general, the trend can be modeled as $\mu(\vecs) = \vecxi(\vecs
)\vecomega$, where
$\vecxi(\vecs) = [\xi_1(\vecs),\ldots,\break \xi_q(\vecs)]$ are fixed
covariate functions and
$\vecomega= [\omega_1,\ldots,\omega_q]^T$ are unknown parameters.
We\vspace*{1pt} found that a linear trend is a plausible assumption and therefore took
$\vecxi(\vecs) = [1,\xi_2(\vecs),\xi_3(\vecs)]$,
$\vecomega= [\omega_1,\omega_2, \omega_3]^T$, where $\xi_2(\vecs)$ and
$\xi_3(\vecs)$ are the latitude and longitude of the location $\vecs$,
respectively. Further, of the various parametric models considered, an
exponential covariance appeared to reflect spatial dependence in the
data best and we therefore assume that
\begin{equation}
\label{expcov}
C_S\bigl(\vecs,\vecs'\bigr) = C(r,
\vecnu)=\cases{ %
\nu_1\exp(-r/\nu_2), & \quad$
\mbox{if } r>0$,\vspace*{3pt}
\cr
\nu_3+\nu_1, & \quad$
\mbox{otherwise}$,}
\end{equation}
where $r$ is the great circle distance between $\vecs$ and $\vecs'$ (in
kilometers), and $\vecnu=[\nu_1,\nu_2,\nu_3]$ with
$\nu_1,\nu_2,\nu_3 \geq0$.

By the construction described in Section~\ref{spat-tempmodel}, the
first diagonal element of the temporal covariance matrix
$\vecC_T$ is equal to 1. It follows that, in the prior model (\ref
{finalprior}) based on the separable covariance (\ref{Kronecker}), the
marginal covariance of the modern temperatures
$[\widetilde{x}_{11}^f, \ldots, \widetilde{x}_{C1}^f]^T$ at the core
lakes is equal to $\vecC_S$. We therefore take $\vecC_S = \vecC_S(\vecnu
)=[C_S(\vecs_c,\vecs_{c'})]$, where $\vecs_c$ and $\vecs_{c'}$ are the
core locations
and $C_S(\vecs_c,\vecs_{c'})$ is computed from (\ref{expcov}).

Denote then by $\vecs_1,\ldots,\vecs_n$ the locations of the training
lakes where the modern mean temperature is known and let
\[
\vecx^m = %
\lleft[ \matrix{ x^m(
\vecs_1)
\cr
\vdots
\cr
x^m(\vecs_n) }
\rright],\qquad \vecxi= %
\lleft[\matrix{ \vecxi(
\vecs_1)
\cr
\vdots
\cr
\vecxi(\vecs_n) } \rright],
\qquad \vecdelta= %
\lleft[\matrix{ \delta(\vecs_1)
\cr
\vdots
\cr
\delta(\vecs_n)} \rright].
\]
Including the spatio-temporal model in the hierarchy,
the formula (\ref{prior2}) is replaced by
\begin{eqnarray}
&& p\bigl(\widetilde{\vecX}^f,\vecvartheta^m,\vecnu,
\vecomega,\kappa|\vecx^m\bigr)\nonumber\\
 &&\qquad= p\bigl(\widetilde{
\vecX}^f|\vecvartheta^m,\vecnu,\vecomega,\kappa,\vecx
^m\bigr)p\bigl(\vecnu,\vecomega|\vecvartheta^m,\kappa,
\vecx^m\bigr)p\bigl(\vecvartheta ^m,\kappa|
\vecx^m\bigr)
\\
\nonumber
&&\qquad= p\bigl(\widetilde{\vecX}^f|\vecnu,\kappa\bigr)p\bigl(
\vecnu,\vecomega|\vecx ^m\bigr)p\bigl(\vecvartheta^m
\bigr)p(\kappa),
\end{eqnarray}
where $p(\vecvartheta^m)$ can be further factored as in (\ref{prior2}).
The first factor on the right-hand side is defined by (\ref
{finalprior}) and the second factor can be further developed as
\[
p\bigl(\vecnu,\vecomega|\vecx^m\bigr) \propto p(\vecnu,\vecomega)p
\bigl(\vecx^m|\vecnu ,\vecomega\bigr).
\]
We assume that, given the parameters $\vecnu$ and $\vecomega$, the
modern temperatures $\vecx^m$ (or the residuals $\vecdelta$) follow a
multivariate normal distribution,
\[
\vecx^m|\vecnu,\vecomega\sim \mathrm{N}\bigl(\vecxi\vecomega,
\vecC_S^m(\vecnu)\bigr),
\]
where $\vecC_S^m(\vecnu) =
[C_S(\vecs_i,\vecs_{j})]$, $\vecs_i$ and $\vecs_{j}$ are training lake locations
and $C_S(\vecs_i,\vecs_{j})$ is computed from (\ref{expcov}).
The prior distributions for parameters $\vecnu$ and $\vecomega$ are
assumed to be independent,
\[
p(\vecnu,\vecomega) = \prod_{i=1}^3 p(
\nu_i)\prod_{i=1}^3 p(
\omega_i).
\]
Considering the notation of Section~\ref{multibummer}, when spatial and
temporal dependence is included in the model, the vector $\vectheta$ of
all model parameters is expanded to
$\vectheta=\{\vecP^m,\vecP^f,\vecvartheta^m,\vecnu,\vecomega,\kappa\}$.

The linear trend in the model is
\begin{equation}
\label{trend} \mu(\vecs) = \vecxi\vecomega= \omega_1 +
\omega_2\xi_2(\vecs) + \omega _3
\xi_3(\vecs),
\end{equation}
where priors for the parameters $\omega_i$ can be elicited by considering
known mean annual temperatures in the part of northern Europe where the
training lakes are located.
Inari in northern Finland ($68^{\circ}$39$'$N, 27$^{\circ}32'$E) and
Tartu in Estonia ($58^{\circ}18'$N, $26^{\circ}44'$E) are located
approximately on the same longitude and the
difference in their annual mean temperatures (years 1981--2010) is
about $-7^{\circ}$C, or about $-0.7^{\circ}$C per a degree of
latitude. We therefore assume that $\omega_2 \sim \mathrm{N}(-1,0.5^2)$.
It is natural to assume that the temperature changes much less in the
east-west direction and, therefore, we take
$\omega_3 \sim\mathrm{N}(0,0.5^2)$. Then, setting $\omega_2 = -0.7$,
$\omega_3 = 0$, and using the fact that the mean annual temperature in
Helsinki ($60^{\circ}10'$N, $24^{\circ}56'$E) is $\mu(\vecs)= 5.9$
$^{\circ}$C, one gets from (\ref{trend}) that $\omega_1 = 47.1^{\circ}$C,
which suggests that a reasonable prior is $\omega_1 \sim \mathrm{N}(47,3^2)$.

Following \citet{TingleyHuybers2010A}, the prior of $\nu_1$ (partial
sill) is an Inverse-gamma distribution,
$\nu_1\sim \operatorname{Inverse\mbox{-}gamma}(0.5,0.2)$, where the parameters were
selected so that the prior is rather vague with mode near a point
estimate of $\nu_1$ (cf. Section~\ref{spat-tempmodel}).
For the range parameter $\nu_2$ we took
$\nu_2\sim\operatorname{Inverse\mbox{-}gaussian}(200,500)$. This conforms to the rule
of thumb suggested in
\citet{Journel1978}, page 194,
since the prior density essentially
vanishes when $\nu_2$ exceeds 800, half the maximum distance over the
field of our data. The nugget parameter $\nu_3$ is assumed to be small,
$\nu_3\sim\operatorname{Gamma}(0.01,10)$.

\subsection{Prior of the temporal smoothing parameter}
\label{EstimationOfKappa}
\begin{figure}

\includegraphics{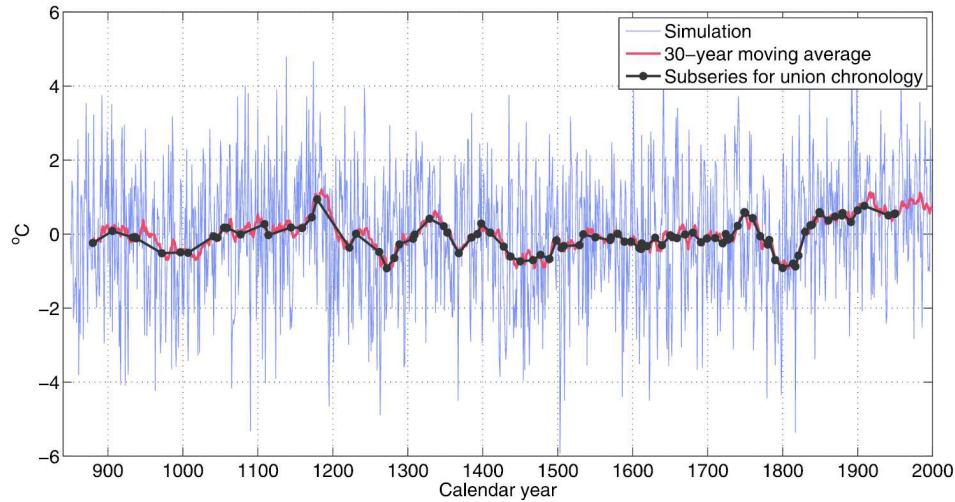}

\caption{Blue curve: NCAR Climate System Model simulation of mean
annual temperature anomaly for the area where the cores are located.
Red curve: 30-year moving average of the simulated anomaly. Black
curve: the 30-year mean evaluated at the union chronology time points.}
\label{Caspar}
\end{figure}

We still need to specify the temporal smoothing parameter $\kappa$ that
encodes our prior beliefs about the variability of past temperatures
$\widetilde{\vecX}^f$. Denoting by $\rho= C_S(\vecs_c,\vecs_c)$ the
diagonal element of the spatial covariance $\vecC_S$ in (\ref{Kronecker}), the marginal prior density of past temperatures at core
$c$ is given by
\begin{equation}
\label{marginalprior}\quad p\bigl(\widetilde{\vecx}_c^f|\vecnu,
\kappa\bigr) = \kappa^{({N-1})/{2}} \exp \Biggl[-\frac{(\widetilde{x}_{c1}^f - \mu
_c)^2}{2\rho}-
\frac{\kappa}{2\rho} \sum_{i=2}^{N} \biggl(
\frac{\widetilde
{x}_{ci}^f-\widetilde{x}_{c(i-1)}^f}{t_{i}-t_{i-1}} \biggr)^2 \Biggr]
\end{equation}
[cf. (\ref{conditionalprior}) and (\ref{properprior})]. If
$\widetilde{\vecx}_c^f$ were known, using a point estimate $\hat{\rho}$
for $\rho$ (cf. Section~\ref{results}), the ``best'' $\kappa$ in the
sense of maximizing (\ref{marginalprior}) would be
\begin{equation}
\label{ML} \hat{\kappa} = \hat{\rho} \Biggl[\frac{1}{N-1} \sum
_{i=2}^{N} \biggl( \frac{\widetilde{x}^f_{ci}-\widetilde
{x}^f_{c(i-1)}}{t_{i}-t_{i-1}}
\biggr)^2 \Biggr]^{-1}.
\end{equation}
In principle, one could try to employ here an existing long
instrumental temperature record but, given that the longest records
cover only the last couple of hundred years,
%(and only about 150 years in case of Finland)
this is not a viable option. Instead of a real instrumental record, we
therefore used an 1150 year long time series of simulated annual mean
temperatures from AD 850 to 1999 for the area where the cores are
located, extracted from the NCAR Climate System Model simulation
described in \citet{Ammetal07}. As the reconstructed temperatures
should be thought of as 30-year annual means (because the training
temperatures are such) at the union chronology time points $t_i$, and
the start of the chronology is commonly taken to be AD 1950, we
restricted the simulated time series to the interval [AD 850, AD 1950],
computed its 30-year moving average from AD 1950 backward, and then
sampled the resulting time series at the times $t_i$. The original
simulation, the moving average and the
subseries corresponding to the union chronology are shown
in Figure~\ref{Caspar}.
%f2

However, besides reconstructions for the union chronology times $t_i$,
we will also be interested in reconstructions for individual core
chronologies (Section~\ref{example}). The problem is that, for some
cores, only a small number of dates between AD 850 and AD 1950
correspond to actual sediment slices (cf. the online supplement),
making estimation of temperature time series roughness dubious. We
therefore extrapolated the roughness information in the simulated time
series to the whole Holocene as follows. From the moving average $z_i$,
$i = 880,\ldots,1999$, we computed for each time difference
$k$ the mean value of $(z_{i} -z_{i-k})^2$ and imputed this value for
$(\widetilde{x}^f_{ci}-\widetilde{x}^f_{c(i-1)})^2$
in (\ref{ML}),\vspace*{1pt} when the interval
$[t_{ci}, t_{c(i-1)}]$ is not contained in the
range [AD 880, AD 1950] and ${t_{ci}-t_{c(i-1)}} = k$. The prior for
$\kappa$ was then defined as
$\operatorname{Gamma}(a,b)$, with $a$ and $b$ selected so that the prior mean
($ab$) is approximately equal to the estimate $\hat{\kappa}$ in
(\ref{ML}) and the prior variance ($ab^2$) is rather large (cf. Table~\ref{kappapriors2}).

%t1
\begin{table}[t]
\tablewidth=250pt
\caption{Parameters of the prior distribution $\operatorname{Gamma}(a,b)$ of the
temperature smoothing parameter $\kappa$ for independent, spatially
independent and spatio-temporal reconstructions, as well as the
corresponding estimate $\hat{\kappa}$ from (\protect\ref{ML})
and the posterior mean of $\kappa$}
\label{kappapriors2}
\begin{tabular*}{250pt}{@{\extracolsep{4in minus 4in}}lccc@{}}
\hline
\textbf{Reconstruction} & $\bolds{a,b}$ & $\hat{\bolds{\kappa}}$ &
$\bolds{\expec(\kappa| \mathrm{data})}$\hspace*{-0.5pt}
\\
\hline
Arapisto & 22, 215 & \phantom{0,}4650 & \phantom{0,}7284\\
Flarken & 412, 49 & 20{,}309 & 20{,}861 \\
Raigastvere & 536, 43 & 23{,}148 & 23{,}313\\
R\~ouge & 19, 230 & \phantom{0,}4343 &10{,}336\\
Union (spatially independent) & 4, 240 & \phantom{0,}1041 &\phantom{0,}9439\\
Union (spatio-temporal) & 1.1, 274 & \phantom{00,}306 &\phantom{0,}4374\\
\hline
\end{tabular*}
\end{table}

%f3
\begin{figure}[b]

\includegraphics{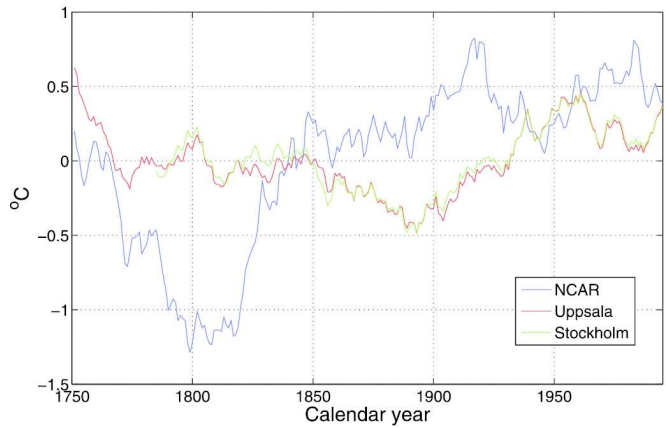}

\caption{30-year moving averages of the computer simulated time series
and instrumental temperature records from Uppsala and Stockholm. All
three records have been centered by subtracting the mean.}
\label{CUS}
\end{figure}

To get an idea how well this procedure might capture the true
characteristics of past temperature variation, we show in Figure~\ref{CUS} the centered 30-year annual means of a part of the simulated
series and two instrumental records, one from Stockholm and one from
Uppsala, two Swedish cities located close to each other and on
approximately the same latitudes as the core lakes used for
reconstruction [\citet{MobergBergstrom1997}]. It appears that the
simulated time series could be a little too rough to mimic actual
temperature variation, at least in the Stockholm--Uppsala area during
the last couple of hundred years. As indicated in Table~\ref{kappapriors2}, the posterior mean of $\kappa$ actually tended to be
larger than the estimate $\hat{\kappa}$ which appears to support this
observation.

\section{Example reconstructions}
\label{example}

\subsection{The data}
\label{data}

Our modern pollen-temperature training set includes $n=173$ lakes with
known 30-year modern annual mean temperatures ($\mu_c$) and surface
sediments analyzed for relative abundances of a total of $l=104$ pollen
taxa. For more details on the training set, see \citet{SeppaEtAl2009}
and \citet{AntonssonEtAl2006}.
Instead of using absolute numbers of pollen grains in sediment samples,
we scaled all counts to the interval $[0,100]$. Although this results in
the loss of some information in the data, the changing environment is
in fact thought to be reflected in the \textit{relative} abundances of
various pollen taxa and not in their absolute numbers. Further,
the absolute total counts at different sites varied greatly (from 169
to 3654) and the Bummer model that underlies our reconstruction methods
appears to work best when the total counts do not differ too much
across sediment samples. A similar scaling of counts to a fixed
interval was also suggested by \citet{HaslettEtAl2006} when, as is
often the case, only the relative abundances pollen taxa are known.

\begin{figure}

\includegraphics{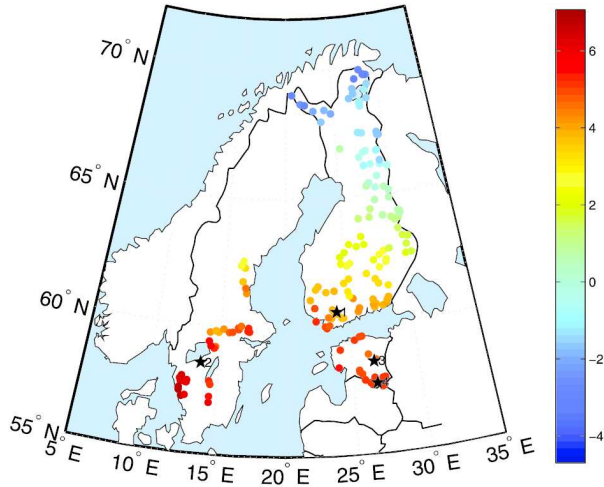}

\caption{Locations of the training lakes and cores. The training lakes are
marked with dots whose colors indicate the associated annual mean
temperature (in ${}^{\circ}$C). The core lakes are as follows: $1 ={}$Arapisto,
$2={}$Flarken, $3 ={}$Raigastvere,
$4 ={}$R\~ouge.}\label{JarvetJaCoretKartalla}
\end{figure}

Past temperature reconstructions were made from four sediment cores
obtained from lakes Arapisto, Flarken, Raigastvere and R\~ouge.
The chronologies of Lakes Arapisto, Flarken and Raigastvere are based
on radiocarbon dating. Conventional bulk radiocarbon datings were
obtained from Flarken (13 datings) and Raigastvere (10 datings) because
these cores were sampled before the use of AMS technique, while the
Arapisto core was dated with 7 AMS datings. All datings were calibrated
and the age--depth curves for all sites were constructed using the
median values of the probability distributions of the calibrated ages.
All three sites have generally stable sedimentation rates, which
increases the reliability of the chronologies [\citet{SeppaPoska2004,SeppaEtAl2005,SarmajaKorjonenSeppa2007}]. Lake R\~ouge
is partly annually laminated but the varve chronology is floating. The
chronology and age--depth model for the lake were derived by correlating
the paleomagnetic secular variation (PSV) curve with the clear anchor
points of the PSV curve of the Finnish varved lake Nautaj\"{a}rvi
[\citet{SeppaEtAl2009}]. The obtained chronology is supported by AMS dates.
Figure~\ref{JarvetJaCoretKartalla} shows the locations of the training
lakes and cores on a map of northern Europe and Table~\ref{table1}
provides additional information on the core lakes.
The four core chronologies consist of a total of 586 time points, but,
as some of these are shared by more than one core, the total number of
dates in the union chronology is only 572
(cf. Table~\ref{table2}). The full chronologies are listed in
\citet{Holetal14suppa}.
For more details, see \citet{SarmajaKorjonenSeppa2007} (Arapisto),
\citet{SeppaEtAl2005} (Flarken), \citet{SeppaPoska2004} (Raigastvere)
and \citet{SeppaEtAl2009} (R\~ouge).

%t2
\begin{table}[b]
\tablewidth=260pt
\caption{The four core lakes used for the pollen-based temperature reconstruction.
The modern temperature is $\mu_c$}
\label{table1}
\begin{tabular*}{260pt}{@{\extracolsep{\fill}}lcccc@{}}
\hline
\textbf{Lake} & \textbf{Latitude} & \textbf{Longitude} & $\bolds{\mu_c}$ \textbf{(${}^{\bolds{\circ}}$C)} &
\textbf{Country} \\
\hline
%\midrule
Arapisto & $60^{\circ}35'$N & $24^{\circ}05'$E & 4.5 & Finland \\
Flarken & $58^{\circ}33'$N & $13^{\circ}40'$E & 5.9 & Sweden \\
Raigastvere & $58^{\circ}35'$N & $26^{\circ}39'$E & 5.0 & Estonia \\
R\~ouge & $57^{\circ}44'$N & $26^{\circ}45'$E & 5.5 & Estonia \\
\hline
\end{tabular*}
\end{table}

%For further information about cores Figure~\ref{JarvetJaCoretKartalla}
%shows the locations of training lakes and cores. Artikkelissa
%\cite{SeppaEtAl2009} sivulla 526 kuvassa 1 n{\chr"E4}kyv{\chr"E4}t
%kaikki 12 FES kalibraatiosetin corea kartalla. Kuten kuvasta n{
%\chr"E4}kyy, coret painottuvat melko etel{\chr"E4}{\chr"E4}n eik{
%\chr"E4} pohjoisesta ole yht{\chr"E4}{\chr"E4}n corea.

\begin{table}
%\tablewidth=280pt
\caption{Details about core chronologies and the union chronology. For each
chronology, shown are its length as well as its youngest and oldest
samples. For the four core lakes, in Section~\protect\ref{notation} these
quantities are denoted by $n_c$, $t_{c1}$ and $t_{cn_c}$, respectively.
Time is expressed as years before present, with 0 corresponding to AD 1950}
\label{table2}
\begin{tabular*}{\tablewidth}{@{\extracolsep{\fill}}lccc@{}}
\hline
\textbf{Core} & \textbf{Chronology length} & \textbf{Youngest sample} & \textbf{Oldest sample} \\
\hline
Arapisto & \phantom{0}98 & \phantom{00}0 & 10{,}852 \\
Flarken & 114 & 118 & 12{,}084 \\
Raigastvere & 115 & \phantom{00}0 & 11{,}594 \\
R\~ouge & 259 & \phantom{00}0 & 11{,}821 \\
Union & 572 & \phantom{00}0 & 12{,}084 \\
\hline
\end{tabular*}
\end{table}

\subsection{The different reconstruction models used}
\label{results}

In addition to the spatio-temporal reconstruction described in
Section~\ref{model}, we also considered two additional approaches.
First, reconstructions were made for each core separately. The model
for each core is exactly the same as in the multi-core case
($C=1$ in Section~\ref{model}),
but with the union chronology replaced by the actual chronology of the
core and the spatial part $\vecC_S$ of
$\vecSigma$ in (\ref{Kronecker}) omitted. We refer to these
reconstructions as ``independent.'' The second variation was to perform
multi-core reconstruction on the union chronology but to replace $\vecC
_S$ by an identity matrix. This ignores distance-based spatial
correlation between the cores but leaves intact interaction through the
shared environmental response parameters $\vecalpha$, $\vecbeta$ and
$\vecgamma$. We refer to this case as ``spatially independent,'' which
refers to a lack of an explicit spatial dependence component in the model.
%The reconstructions based on this model are presented in the on-line
%supplement to this paper.

The prior, the estimate $\hat{\kappa}$ from (\ref{ML}) and the
posterior mean for the smoothing parameter $\kappa$ in each case is
given in Table~\ref{kappapriors2}. In (\ref{marginalprior}), for the
independent and spatially independent models, $\rho=1$, and for the
spatio-temporal model, we took
$\hat{\rho} = 0.2937$, the value obtained from a point estimate of the
spatial covariance (cf. Section~\ref{spat-tempmodel}). We note that
in some cases the posterior mean of $\kappa$ lies far in the right tail
of the prior distribution. We therefore recomputed the reconstruction
in these cases with vague $\kappa$ priors centered at the posterior
means of Table~\ref{kappapriors2}. Now the new posterior means were
quite close to the prior means and the reconstructions themselves
changed little. We therefore believe that the priors of Table~\ref{kappapriors2} are reasonable and result in reliable temperature
reconstructions.

%({\bf comment the difference between prior and posterior means? Yes:
%say that we reran the reconstructions putting the kappa prior means to
%their posterior values using quite vague priors. The reconstructions
%were practically unchanged and now kappa stayed within the priors. See
%Liisa's mail July 15. Further, al in all, reconstructions are very
%insensitive to changing kappa value. And, the role of the prior (vague
%Amman) can be defended: it guards against unreasonable values (like
%too big values)!}).

%\begin{table}[h]
%\caption{Values for smoothing parameter $\kappa_c$ for four cores
%using core's own chronology and union chronology. Time 0 is year 1950.}
%\label{smoothingvalues2}
%\begin{tabular}{lcccc}
%\hline
%name & $n_c^*$ & $\hat{\kappa}_c$ (short) & $n_c$ & $\hat{\kappa}_c$
%(long) \\ \hline
%Arapisto & 5 & 154580 & 98 & 3091 \\
%Flarken & 5 & 459410 & 114 & 20309 \\
%Raigastvere & 11 & 27109 & 115 & 23148 \\
%Rouge & 60 & 1817 & 259 & 4343 \\
%union (independent) & 78 & 1431 & 572 & 1084 \\
%union (spatio-temporal)& 78 & 555 & 572 & 420 \\ \hline
%\end{tabular}
%\end{table}

The posterior means for the parameters of the spatial covariance were
\begin{eqnarray*}
\expec(\vecnu| \mathrm{data}) & =& [0.2108,147.9279, 0.0698]^T ,
\\
\expec(\vecomega| \mathrm{data}) & =& [47.4144, -0.7014, -0.0472]^T.
\end{eqnarray*}
The posterior mean covariance
matrix is
\[
\expec(\vecC_S | \mathrm{data}) = %
\lleft[\matrix{ 0.281
& 0.003 & 0.035 & 0.020
\cr
0.003 & 0.281 & 0.001 & 0.001
\cr
0.035 & 0.001 &
0.281 & 0.111
\cr
0.020 & 0.001 & 0.111 & 0.281} \rright],
\]
where the lakes appear in the order Arapisto, Flarken, Raigastvere and
R\~ouge. Thus, the elements in the first row from left to right show
the variance for Arapisto, the covariance between Arapisto and Flarken,
the covariance between Arapisto and Raigastvere, and the covariance
between Arapisto and R\~ouge, and so on.

For each time point $t_i$, the posterior mean temperature and its $95\%
$ highest posterior density interval were computed. As such point-wise
credible intervals may underestimate the uncertainty in the
paleotemperature time series regarded as a whole curve, we also
calculated a $95\%$ simultaneous credible band employing the method of
``Simultaneous Credible Intervals'' suggested in
\citet{ErastoHolmstrom2005}. Using the generated posterior sample, the method
first finds a $\Delta> 0$ such that
\[
\PP\biggl(\max_{i=1,\ldots,N} \biggl\llvert \frac{\widetilde{x}_{ci}^f-\expec(\widetilde{x}_{ci}^f |\mathrm{data})}{\operatorname{Std}(\widetilde{x}_{ci}^f |
\mathrm{data})} \biggr
\rrvert \leq \Delta \vert \mathrm{data} \biggr) = 0.95
\]
and then defines the simultaneous credible band as
\[
%\left[\expec(\tilde{x}_{ci}^f |\mathrm{data}) - \Delta\operatorname{Std}(
%\tilde{x}_{ci}^f | \mathrm{data}),
%\expec(\tilde{x}_{ci}^f |\mathrm{data}) + \Delta\operatorname{Std}(
%\tilde{x}_{ci}^f | \mathrm{data})\right],
%\;\; i = 1,\ldots,N.
\expec\bigl(
\widetilde{x}_{ci}^f |\mathrm{data}\bigr) \pm\Delta
\operatorname{Std}\bigl(\widetilde {x}_{ci}^f |
\mathrm{data}\bigr),\qquad i = 1,\ldots,N.
\]
The point-wise and simultaneous credible intervals are probability
intervals based on the posterior probability which itself is determined
by the data and the model assumptions. The reconstruction accuracy of a
simplified version of our single-lake model (essentially the Bummer
model) was checked using training set cross-validation in
\citet{VaskoEtAl2000,ToivonenEtAl2001} and \citet{SalonenEtAl2012},
where it was found that, in terms of root mean square error of
prediction, it performed competitively against standard methods, such
as WA-PLS. The structure of the spatio-temporal prior prevents such
validation for the more complex model considered here.

\subsection{The Gaussian response model}
The plausibility of the Gaussian response model of Section~\ref{multibummer} is discussed extensively in the online supplement
[\citet{Holetal14suppa}] and we summarize here the main conclusions. First,
based on comparisons with the training data, the model appears to
describe the observed relative taxon abundances reasonably well. The
overall character of predicted abundances as a function of temperature
also seem plausible with nearby lakes and the multi-core
reconstructions producing similar response curves. For most taxa the
optimal temperature ranges suggested by the estimated response curves
do not seem unreasonable. The similarity of the response curves of the
two multi-core reconstructions (spatially independent and
spatio-temporal models) is consistent with the similarity of their
temperature reconstructions (see below). In the case of the more
southern lakes (Flarken, Raigastvere and R\~ouge), the estimated peak
relative abundances of warmer temperature taxa exceeds the abundance
seen in the training and core data, while for the northernmost lake
(Arapisto), these abundances are considerably lower. This is not
unexpected, considering the modest share of most of these warmer
temperature taxa in the Arapisto core.

The posterior values of the optimal temperature parameter $\beta_j$ for
the warmer temperature taxa substantially exceed their prior values.
This may be explained by the fact that the prior is centered on the
optimal value estimated from the training set and the training lake
temperatures are likely to be considerably lower than many of the past
temperatures at the core lakes. The posterior mean of $\beta_j$ only
roughly corresponds to the temperature at which the modeled abundance
probability of taxon $j$ peaks, although this correspondence seems to
be more robust for the multi-core reconstructions.
Therefore, one should not interpret the parameter $\beta_j$ as
representing a precise optimal taxon temperature. Also, the posterior
values of the tolerance parameter $\gamma_j$ tend to be very large,
making the Gaussian response function [(\ref{modernlambda}) and (\ref
{fossillambda})] flat, undermining $\beta_j$'s role as a clearly
defined optimum temperature.

\subsection{Interpretation of reconstructed temperature histories}

\begin{figure}

\includegraphics{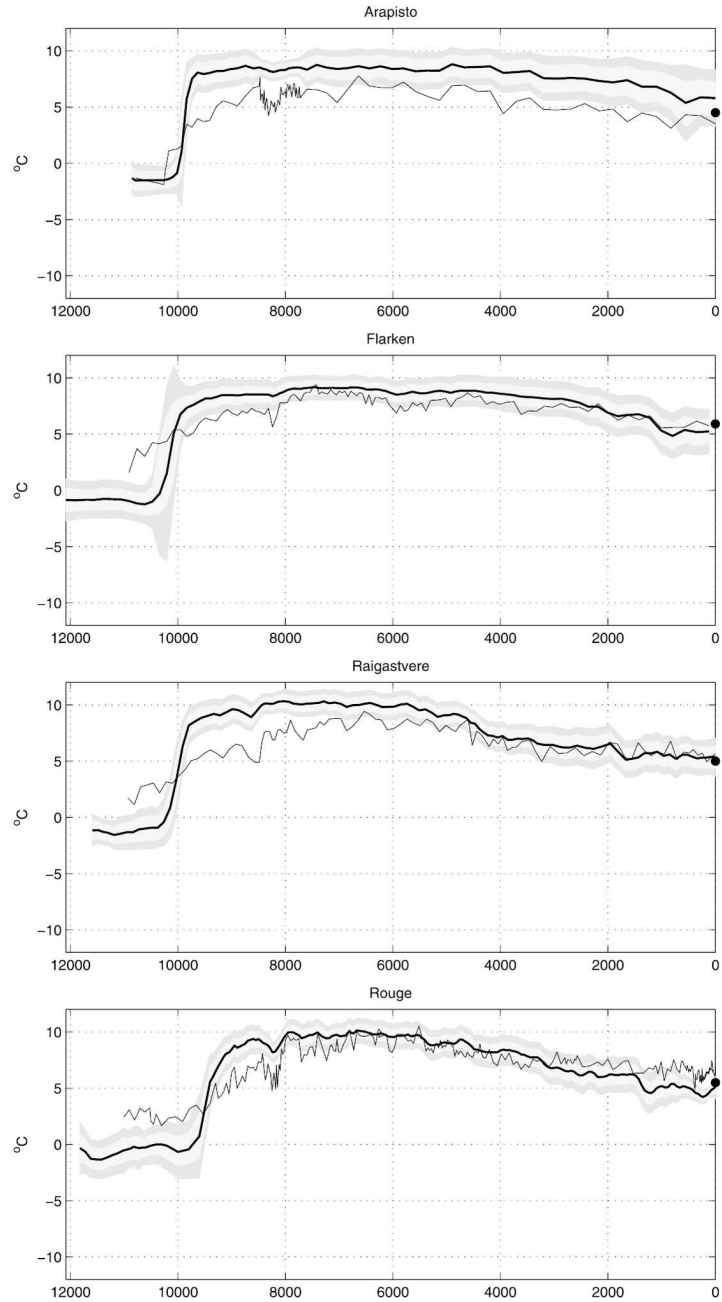}
\vspace*{-4pt}
\caption{Temperature reconstructions made independently from each core.
The thick curve is the posterior mean and the thin curve is the WA-PLS
reconstruction.
Light and dark gray show the point-wise and
simultaneous 95\% credible bands, respectively. Horizontal axis: time
in years before present.
Vertical axis: mean annual temperature in centigrades. The dot at AD
1950 marks the current mean annual instrumental temperature.}
\label{singlecores}
\end{figure}

\begin{figure}

\includegraphics{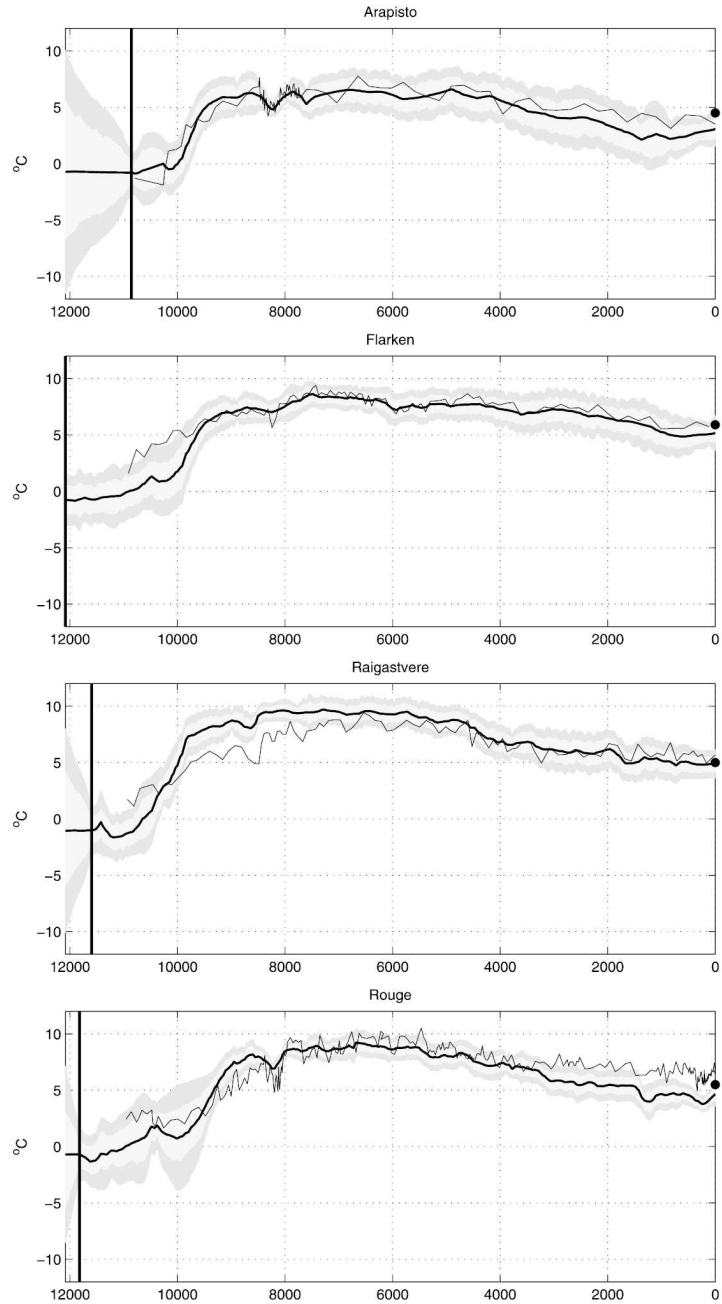}
\vspace*{-4pt}
\caption{Temperature reconstructions based on the spatially independent
model with no explicit spatial interaction.
The thick curve is the posterior mean and the thin curve is the WA-PLS
reconstruction.
Light and dark gray show
the point-wise and simultaneous 95\% credible bands, respectively. For
each lake, the black line marks the oldest date in its own chronology.
Horizontal axis: time in years before present.
Vertical axis: mean annual temperature in centigrades. The dot at AD
1950 marks the current mean annual instrumental temperature.}
\label{independent}
\end{figure}

\begin{figure}

\includegraphics{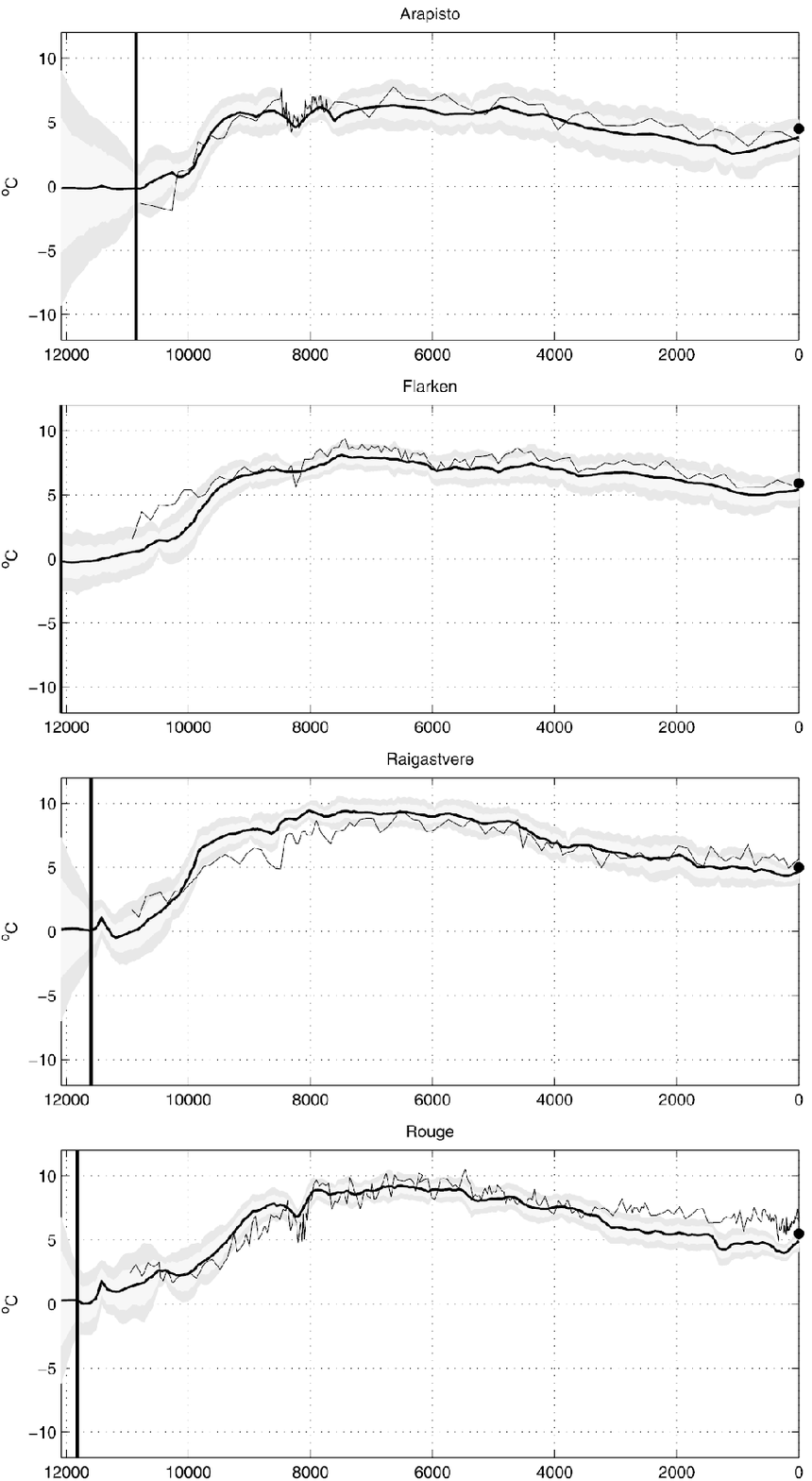}
\vspace*{-4pt}
\caption{Temperature reconstructions based on the full spatio-temporal
model.
The thick curve is the posterior mean and the thin curve is the WA-PLS
reconstruction.
 Light and dark gray show the point-wise and simultaneous 95\%
credible bands, respectively. For each lake, the black line marks the
oldest date in its own chronology. Horizontal axis: time in years
before present.
Vertical axis: mean annual temperature in centigrades. The dot at AD
1950 marks the current mean annual instrumental temperature.}
\label{spatiotemporal}
\end{figure}

The past temperature reconstructions for different models are shown in
Figures~\ref{singlecores}--\ref{spatiotemporal}. For comparison, we also include in all figures
reconstructions made with the WA-PLS method [\citet{terBraakJuggins1993}], one of the most popular calibration methods
used in pollen-based reconstructions. The source for these
reconstructions was \citet{SeppaEtAl2009}.

Figure~\ref{singlecores} displays the independent reconstructions for
each lake.
Figure~\ref{independent} shows the spatially independent
reconstructions and Figure~\ref{spatiotemporal} shows the
reconstructions made with the full spatio-temporal model.
In all figures, the thick curve is the posterior mean, the thin curve
is the WA-PLS reconstruction and a dot at AD 1950 marks the current
mean annual instrumental temperature. Lighter and darker gray show the
point-wise and simultaneous 95\% credible bands, respectively.
In Figures~\ref{independent} and \ref{spatiotemporal}, for each lake,
the black line marks the oldest date in its own chronology.

Our first observation is that for each lake the general features of the
reconstructions based on the spatially independent and spatio-temporal
models are quite similar and both differ to some extent from the
reconstructions made independently from single cores. Allowing the
cores to interact, either through shared parameters or spatial
correlation, also makes reconstructions for all lakes more similar.
Further, the full spatio-temporal reconstructions generally have least
posterior uncertainty, as exhibited by the smaller credibility
intervals. %especially in the early part of the Holocene.
From roughly 7700 years before present onward, the uncertainties are
% in the independent and spatio-temporal reconstructions are generally
%smallest in the R\~ouge record, which has the highest sample
%resolution of the four sites, and
largest in the Arapisto record, which also has the lowest sediment
sample resolution for this period (cf. the online supplement).
The large uncertainties to the left of the black lines in Figures~\ref{independent} and \ref{spatiotemporal} are due to lack of pollen data
for the lake in question, which causes the reconstructed temperatures
to be supported only by the priors.
%f5
%f6
%f7

The distinct feature in the results obtained with independent
reconstructions from single cores is the abrupt rise of temperature
during the early Holocene. These single core reconstructions also show
generally higher temperature values during the mid-Holocene (roughly
8000--4000 years ago) than in the WA-PLS-based reconstructions
(Figure~\ref{singlecores}). In the spatially independent
reconstructions, the temperature values during the mid-Holocene are
more consistent with those generated with the WA-PLS technique
(Figure~\ref{independent}). The reconstructions based on the full
spatio-temporal model (Figure~\ref{spatiotemporal}) show trends which
are most compatible with the WA-PLS-based trends, with most gradual
temperature rise in the early Holocene. The Holocene thermal maximum,
the warmest period of the Holocene, strongly expressed in northern
Europe in paleoclimatic data and model simulations [\citeauthor{Renetal09} (\citeyear{Renetal09,Renetal12})], is observable in the spatially independent and
full spatio-temporal model reconstructions at about 8000--5000 years
ago, with temperature value of
8--9$^{\circ}$C at the R\~ouge, Raigastvere and Flarken, and about
6$^{\circ}$C at the northernmost site Arapisto in Finland. These
patterns are generally concordant with the WA-PLS results. Moreover,
the results show that the high sample resolution helps decrease the
uncertainty in the reconstructions. This is reflected particularly in
the Arapisto and R\~ouge records, where a higher number of pollen
samples were analyzed between 8500--8000 years ago to detect possible
indications of an abrupt cold event widely observed in northern Europe
[\citet{AllAgu05,WieRen06}]. This event is reflected in R\~ouge and
Arapisto data by a $\sim$1$^{\circ}$C temperature dip, while the
influence of high sample resolution is apparent by markedly smaller
statistical uncertainties at 8000--8500 years ago (Figure~\ref{spatiotemporal}). The cold event 8200 years ago is present also in the
WA-PLS based reconstructions from Flarken, but it shows less clearly in
the posterior means, presumably due to low temporal sample resolution
of this record. The Little Ice Age (about AD 1550 to 1850) and the
subsequent warming show best in the spatio-temporal reconstructions.

Two features in these reconstructions require further analysis. The
first is the rate and magnitude of early Holocene warming that appears
quite different for the independent reconstructions (Figure~\ref{singlecores}) and the joint, multi-core reconstructions (Figures~\ref{independent} and \ref{spatiotemporal}). The second question concerns
timing of the onset of warming.

An online supplement [\citet{Holetal14suppa}] includes published
reconstructions from Greenland ice cores that often are used as a
reference when Holocene
climate is reconstructed for Northern Europe and the North Atlantic
region (Figures~S.5 and S.6).
The reconstruction in Figure~S.6 suggests that the early rise in
temperature has been
6--10$^{\circ}$C (depending on the amount of smoothing applied) and
our reconstructions are within that range. However, according to the
ice core records, the rate at which the temperature rises in the
individual, single-core reconstructions is too high. This view is also
supported by the Scandinavian reference reconstructions (Figures~S.7,
S.8 and S.9 in the supplement) as well as the WA-PLS reconstructions
displayed in Figures~\ref{singlecores}, \ref{independent} and \ref
{spatiotemporal}. The global and hemispheric reconstructions in
\citet{Mar2013} and \citet{Sha2012} also support this conclusion, although
they may be less relevant than the more local Greenland and
Scandinavian records. Looking at the pollen abundances in the four
cores (Figure~S.10), we notice that \textit{Alnus} (alder), \textit{Corylus}
(hazel), \textit{Ulmus} (elm) and \textit{Tilia} (linden) are among the taxa
whose growing abundance coincides with the onset of warming. These are
all taxa with optimal temperatures that are likely to be higher than
the past temperatures at the four core lakes (see also Figure~S.2 in
the online supplement). This can explain the timing of the temperature
rise reconstructed for these lakes, but it does not clarify why the
single-core Bayesian model seems to overestimate the rate of
temperature change. If desired, this could be remedied by increasing
considerably the value of the temporal smoothing parameter $\kappa$ in
(\ref{depmodel}), but such an ad hoc choice might be difficult to
justify. Alternatively, one might use a time-dependent smoothing
parameter that would smooth the onset of warming differently from the
rest of the Holocene. Similarly to our present approach, such a choice
could perhaps be based on numerical climate simulations. Such
considerations are left for future research.

Compared with the single-core reconstructions, the rate of warming in
the spatially independent and spatiotemporal joint reconstructions is
in much better agreement with the Greenland and Scandinavian reference
records as well as the WA-PLS reconstructions. It appears that the
potential difficulty the single-core Bayesian model has in handling
such a rapid rise can be alleviated by borrowing strength from other
cores. Thus, sharing the abundance model parameters between the cores,
and therefore effectively increasing the number of data available for
their estimation, already makes a significant difference. Spatial
smoothing then further tempers the reconstructed temperature rise.

One might, however, suspect that the apparent difference in the timing
of the onset of warming in the four independent reconstructions alone
when combined with correlations between the reconstructions explains
the more gradual warming in the joint reconstructions. Indeed, while
Arapisto, Flarken and Raigastvere temperatures start to rise almost
simultaneously, the onset of warming for R\~ouge appears to take place
later (Figure~\ref{singlecores}). This is all the more problematic
since Lake R\~ouge is the southernmost of the four core lakes, and
therefore would be expected to warm first. Such a discrepancy could be
explained by the rather wide confidence intervals around 10{,}000 BP, but
another possibility is the relative paucity of chronology dates for
Lake R\~ouge between 10{,}200 BP and 9400 BP (cf. the online supplement).
We therefore made reconstructions also with R\~ouge data before 9400 BP
left out. The results are shown in Figures~S.10, S.11 and S.12 in the
online supplement. While the temperature rise in the joint
reconstructions is now somewhat sharper than in Figures~\ref{independent} and \ref{spatiotemporal}, it is still much more gradual
than in the single-core reconstructions of Figure~\ref{singlecores}.
We conclude that the main factor in decreasing the rate of early
Holocene warming in the joint reconstructions is sharing of the
taxon-specific response parameters. This, of course, does not exclude
the possibility of additional smoothing in the joint reconstructions
because of chronology misalignments caused by dating errors. The error
in the radiocarbon dates varies between the four lakes and depends on
the age of the sediment sample, being generally larger for the oldest
samples. Thus, errors of 100--200 years are likely for the oldest
samples of lakes Arapisto, Flarken and Raigastvere, but for Lake R\~
ouge they can be even larger. Even with the earliest R\~ouge data left
out, some smoothing may therefore result at the time of early Holocene
warming because the reconstructions may not be correctly aligned.
The best solution would be to let the dating errors influence the
reconstructions and their posterior uncertainty by
incorporating them in the hierarchical model. A simple additive error
model was proposed in
\citet{ErastoEtAl2012}, but a more satisfactory approach would include
a sophisticated Bayesian chronology model such as the Bchron of
\citet{HasPar08} as a model component. We will consider this in future work.

\subsection{Computational details}

In all cases, a Metropolis-within-Gibbs sampler [e.g., \citet{RobertCasella}] was run for 30{,}000 iterations, the first 15{,}000 were
used for burn-in and from the last 15{,}000, every 5th sample was kept for
inference. Thus, each posterior analysis was based on a sample of size
3000. The relevant conditional posterior distributions are given in
Appendix~\ref{conditionals}.

%In all cases, a sample of 30 000 posterior realizations of the
%paleotemperature $\widetilde{\vecx}_c^f$ (for individual-core
%reconstruction) or $\widetilde{\vecX}_c^f$ (for multi-core
%reconstruction) was generated using Metropolis-within-Gibbs sampling
%(e.g.\ \cite{RobertCasella}). The relevant conditional posterior
%distributions are given in Appendix~\ref{conditionals}.

In both spatially independent and spatio-temporal reconstructions, some
chronology time points are not associated with corresponding pollen
abundance data. Our strategy was to first update, one by one, the
temperatures which do have associated pollen data and after that those
without pollen data, conditioning them on those with pollen data.
Adaptive simulation was used both for temperatures and the
environmental response parameters [\citet{Gelman}]. The adaptive phase
consisted of 10{,}000 iterations and the subsequent fixed phase of 20{,}000
iterations that used the proposal variances from the last adaptive step.

The initial values for the components of the temperature vector
$\widetilde{x}_c^f$ were simulated from $\mathrm{N}(\mu_c,1.5^2)$, where $\mu_c$
is the modern temperature at core $c$ (see Table~\ref{table1}). The
initial values of $\kappa$, the trend parameters $\omega_i$ and the
range parameter
$\nu_2$ were generated from their priors. For partial sill $\nu_1$, the
mode of the prior was used and the nugget $\nu_3$ was initialized at
its prior mean.

The abundances among the taxa analyzed vary considerably, with many
taxa appearing in the sediment samples only rarely and only some
appearing in substantial abundance. We therefore thought it best to use
taxon-specific initialization for the scaling factor $\alpha_j$.
The iteration for $\alpha_j$ was started at $\alpha_{j,\mathrm{max}}/2$, where
$\alpha_{j,\max}$ is the largest observed abundance of taxon $j$.
Such initialization accelerated convergence substantially. The initial
values for $\beta_j$ and $\gamma_j$ were generated from their prior
distributions $\mathrm{N}(\hat{\beta}_j,(1.5\sqrt(3))^2)$ and $\operatorname{Gamma}(9,1/3)$,
respectively.

The algorithms were implemented in Matlab and run on a PC with an Intel
Core i7 3770 CPU.
Table~\ref{times} summarizes approximate computation times in different cases.

%t4
\begin{table}
\tablewidth=220pt
\caption{CPU times for temperature reconstructions. For each lake, the CPU time
is for reconstruction made independently using its own core chronology.
The last two reconstructions are joint reconstructions using all four cores}
\label{times}
\begin{tabular*}{220pt}{@{\extracolsep{\fill}}lc@{}}
\hline
\textbf{Reconstruction} & \textbf{CPU time (hours)} \\
\hline
Arapisto & \phantom{0}3 \\
Flarken & \phantom{0}3 \\
Raigastvere & \phantom{0}3 \\
R\~ouge & \phantom{0}3 \\
Union (spatially independent) & 26 \\
Union (spatio-temporal) & 28\\
\hline
\end{tabular*}
\end{table}

\section{Conclusions}
\label{discussion}

We propose a novel Bayesian approach for the reconstruction of past
temperature variation during the Holocene, using fossil pollen data
from multiple sediment cores.
A spatio-temporal model was described that takes into account both
temporal correlations within the cores and spatial correlations between
them. Temporal correlations were modeled with a smoothing prior where
the smoothing parameter hyperprior was elicited using numerical climate
simulation. The temporal smoothing prior is very vague and favors
rather slowly varying temperature time series, which is consistent with
the relatively stationary climate conditions during the Holocene.
An isotropic covariance was used to model the spatial dependence of the
temperatures across
the sites from which the sediment samples were obtained.

Taking into account spatial dependencies between reconstructions
reduced uncertainty and made their overall shapes more similar. Given
that the four cores considered are from a geographically restricted area
and that the temperature history at the four sites therefore must have
been similar, it can be argued that the spatio-temporal reconstructions
are an improvement over the reconstructions made independently from
each core or those without explicit spatial dependencies. The
spatio-temporal reconstructions are also smoother, less uncertain and
generally more realistic.
% than the single core reconstructions.
In addition, they are more consistent with the results obtained with
WA-PLS, a popular method for pollen-based reconstructions.

The proposed model is directly applicable to reconstructions from other
 biological proxies records, such as diatoms and chironomids. Other climate variables
besides temperature could also be considered.\vadjust{\goodbreak} It would also be
interesting to consider a larger set of proxy records from a more
extensive geographic area. In some situations a nonstationary spatial
covariance might have to be used to model different types of
correlations within and between distinctly different types of regions.
Finally, the chronologies were assumed error-free, which of course is a
simplification. Therefore, future work will need to also address the
uncertainty related to the various sources of errors involved in
constructing the chronologies.

\begin{appendix}
\section{Temporal covariance structure}
\label{algebra}

Computation of the covariance matrix $\vecGamma$ in (\ref{properprior})
is needed for efficient
implementation of the sampling procedures used in estimation.
Following, for example,   \citet{KaipioSomersalo}, the quadratic form
in the exponent of
(\ref{conditionalprior}) is first written as
\[
\sum_{i=2}^{N} \biggl(\frac{\widetilde{x}_{ci}^f-\widetilde
{x}_{c(i-1)}^f}{t_{ci}-t_{c(i-1)}}
\biggr)^2= \bigl\llVert \vecL\widetilde{\vecx}_c^f
\bigr\rrVert ^2,
\]
where\vspace*{-2pt} $\vecL= [L_{ij}] \in\reals^{(N-1)\times N}$,
\[
L_{ij} = \cases{ -(t_{c(i+1)}-t_{ci})^{-1},
& $\quad\mbox{when $j=i$}$, \vspace*{1.5pt}
\cr
(t_{c(i+1)}-t_{ci})^{-1},
& $\quad\mbox{when $j=i+1$}$,\vspace*{1.5pt}
\cr
0, & $\quad\mbox{otherwise}$.}
\]
Let\vspace*{-3pt}
\[
\vecL^T\vecL= %
\lleft[\matrix{ \vecB_{11} &
\vecB_{12}
\cr
\vecB_{21} & \vecB_{22}} \rright],
\]
where $\vecB_{11}\in\reals$, $\vecB_{12}\in\reals^{N-1}$, $\vecB_{21}\in
\reals^{(N-1)\times1}$ and $\vecB_{22}\in\reals^{(N-1)\times(N-1)}$.\vspace*{-2pt}
Then
\[
p\bigl(\widetilde{\vecx}_{c*}^f |\widetilde{x}_{c1}^f,
\kappa\bigr) \propto\kappa ^{({N-1})/{2}} \exp \biggl[-\frac{\kappa}{2} \bigl(
\widetilde{ \vecx}_{c*}^f+\vecB _{22}^{-1}
\vecB_{21}\widetilde{x}_{c1}^f
\bigr)^T \vecB_{22} \bigl( \widetilde{\vecx}_{c*}^f+
\vecB_{22}^{-1}\vecB_{21}\widetilde{x}_{c1}^f
\bigr) \biggr],
\]
where $\vecB_{22}^{-1}\vecB_{21}\widetilde{x}_{c1}^f=[\widetilde
{x}_{c1}^f,\ldots, \widetilde{x}_{c1}^f] \in\reals^{N-1}$. The formula\vspace*{1pt}
(\ref{properprior}) then follows readily when
$\widetilde{x}_{c1}^f \sim\mathrm{N}(\mu_c,1)$.\vspace*{-2pt}

\section{The conditional posteriors}
\label{conditionals}

From (\ref{condindep})--(\ref{dub}) and\vspace*{-1pt} Section~\ref{spat-tempmodel},
%The posterior distribution $p(\vecX^f,\vectheta|\vecY^f,\vecx^m,
%\vecY^m)$ can be obtained using (\ref{condindep}), (\ref{multin}), (
%\ref{prior}), (\ref{Dirmod}), (\ref{dub}), (\ref{prior2}) and (
%\ref{endprior}). The result is
\begin{eqnarray*}
&& p\bigl(\widetilde{\vecX}^f, \vectheta|\vecY^f,
\vecx^m,\vecY^m\bigr)
\\[-2pt]
&&\qquad \propto \prod_{c=1}^C \prod
_{i=1}^{n_c}p\bigl(\vecy ^f_{ci}|x^f_{ci},
\vectheta\bigr) \prod_{i=1}^np\bigl(
\vecy^m_i | x^m_i,\vectheta
\bigr) \prod_{i=1}^n p\bigl(
\vecp^m_i | x^m_i,
\vecvartheta^m\bigr)
\\[-2pt]
&&\qquad\quad {}\times\prod_{c=1}^C
\prod_{i=1}^{n_c} p\bigl(\vecp^f_{ci}
| x^f_{ci},\vecvartheta ^f_c\bigr) p\bigl(\widetilde{\vecX}^f|\vecnu,\kappa,
\bigr)p(\kappa) p\bigl(\vecx^m|\vecnu,\vecomega\bigr)\prod
_{i=1}^3 p(\omega_i)\prod
_{i=1}^3 p(\nu_i) \\[-2pt]
&&\qquad\quad {}\times\prod
_{j=1}^l p(\alpha_j) \prod
_{j=1}^l p\bigl(\beta_j |
\vecx^m\bigr) \prod_{j=1}^l p(
\gamma_j)
\\[-2pt]
&&\qquad=\prod_{c=1}^C \prod
_{i=1}^{n_c} \operatorname{Mult}\bigl(
\vecy^f_{ci}|y^f_{ci\cdot
},
\vecp^f_{ci}\bigr) \prod_{i=1}^n
\operatorname{Mult}\bigl(\vecy^m_{i}|y^m_{i\cdot},
\vecp^m_i\bigr)
\\[-2pt]
&&\qquad\quad {}\times \prod_{i=1}^n
\operatorname{Dirichlet}\bigl(\vecp_i^m |
\veclambda_i^m\bigr) \prod_{c=1}^C
\prod_{i=1}^{n_c}\operatorname{Dirichlet}
\bigl(\vecp_{ci}^f | \veclambda_{ci}^f
\bigr)
\\[-2pt]
&&\qquad\quad {}\times\mathrm{N}\bigl(\widetilde{\vecX}^f|\vecmu,
\vecSigma\bigr)\times\operatorname{Gamma}(\kappa| a,b)\times \mathrm{N}\bigl(
\vecx^m|\vecxi\vecomega,\vecC_S^m(\vecnu)
\bigr)\times\mathrm{N}(\vecomega|\vecmu_{\vecomega_0},\vecSigma_{\vecomega_0})
\\[-2pt]
&&\qquad\quad {}\times\operatorname{Inv\mbox{-}gamma}(\nu_1|0.5,0.2)\times
\operatorname{Inv\mbox{-}gaussian}(\nu_2|200,500)\\
&&\qquad\quad {}\times\operatorname{Gamma}(
\nu_3|0.01,10)
\\[-2pt]
&&\qquad\quad {}\times\prod_{j=1}^l
\operatorname{Unif}(\alpha_j | 0.1,50) \prod
_{j=1}^l \mathrm{N}\bigl(\beta_j |
\hat{\beta}_j,(1.5\sqrt{3})^2\bigr) \prod
_{j=1}^l\operatorname{Gamma}(\gamma_j |
9,1/3),
\end{eqnarray*}
where $\vecSigma=\vecC_S(\vecnu) \otimes\vecC_T(\kappa)$, $\vecmu
_{\vecomega_0}=[47,-1,0]$, and
$\vecSigma_{\vecomega_0}=\operatorname{diag}(3^2,0.5^2,0.5^2)$.
%\begin{bmatrix}
%3^2 & 0 & 0 \\
%0 & 0.5^2 & 0 \\
%0 & 0 & 0.5^2 \\
%\end{bmatrix},
%$

Therefore, the full conditional posterior distributions of the unknown
parameters are
%
%The full conditional posterior of each unknown parameter can now be
%obtained and they are given below. The notation $X|\cdot$ indicate the
%conditional distribution of $X$ given parameters and data.
\begin{eqnarray*}
p\bigl(\widetilde{\vecX}^f|\cdot\bigr) &\propto & \prod
_{c=1}^C\prod_{i=1}^{n_c}
\operatorname{Dirichlet}\bigl(\vecp^f_{ci} |
\veclambda^f_{ci}\bigr) \exp \biggl[-\frac{1}{2}\bigl(
\widetilde{\vecX}^f-\vecmu\bigr)^T\vecSigma
^{-1}\bigl(\widetilde{\vecX}^f-\vecmu\bigr) \biggr],
\\
p(\kappa|\cdot) &\propto & \kappa^{ ({C(N-1)+2(a-1)})/{2}}\\
&&{}\times \exp \biggl[-\frac{\kappa}{b}
- \frac{1}{2}\bigl(\widetilde{\vecX}^f-\vecmu
\bigr)^T\bigl(\vecC_S(\vecnu) \otimes\vecC_T(
\kappa)\bigr)^{-1}\bigl(\widetilde{\vecX}^f-\vecmu\bigr)
\biggr],
\\
p\bigl(\vecp_i^m | \cdot\bigr) & = &
\operatorname{Dirichlet}\bigl( \vecp_i^m |
\vecy_i^m+\veclambda_i^m\bigr),
\\
p\bigl(\vecp_{ci}^f | \cdot\bigr) &=&
\operatorname{Dirichlet}\bigl(\vecp_{ci}^f | \vecy
_{ci}^f+\veclambda_{ci}^f\bigr),
\\
p(\alpha_j|\cdot) &\propto & \prod_{i=1}^n
\operatorname{Dirichlet}\bigl(\vecp^m_i |
\veclambda^m_i\bigr) \prod_{c=1}^C
\prod_{i=1}^{n_c} \operatorname{Dirichlet}
\bigl(\vecp^f_{ci} | \veclambda^f_{ci}
\bigr) \times \operatorname{Unif}(\alpha_j | 0.1,50),
\\
p(\beta_j|\cdot) &\propto & \prod_{i=1}^n
\operatorname{Dirichlet}\bigl(\vecp^m_i |
\veclambda^m_i\bigr) \prod_{c=1}^C
\prod_{i=1}^{n_c} \operatorname{Dirichlet}
\bigl(\vecp^f_{ci} | \veclambda^f_{ci}
\bigr) \times \mathrm{N}\bigl(\beta_j | \hat{\beta}_j,
(1.5\sqrt{3})^2\bigr),
\\
p(\gamma_j |\cdot) &\propto & \prod_{i=1}^n
\operatorname{Dirichlet}\bigl(\vecp^m_i |
\veclambda^m_i\bigr) \prod_{c=1}^C
\prod_{i=1}^{n_c} \operatorname{Dirichlet}
\bigl(\vecp^f_{ci} | \veclambda^f_{ci}
\bigr) \times \operatorname{Gamma}(\gamma_j | 9,1/3),
\\
p(\vecomega|\cdot)&=&\mathrm{N}\bigl(\vecomega| \vecSigma_{\vecomega}\bigl(
\vecxi^T\vecC^m_S(\vecnu)^{-1}
\vecx^m+ {\vecSigma}_{\vecomega_0}^{-1}
\vecmu_{\vecomega_0} \bigr),{\vecSigma}_\vecomega\bigr),
\end{eqnarray*}
where
\[
\vecSigma_\vecomega = \bigl(\vecxi^T\vecC^m_S(
\vecnu)^{-1}\vecxi+ \vecSigma _{\vecomega_0}^{-1}
\bigr)^{-1},
\]
\begin{eqnarray*}
p(\nu_i|\cdot) & \propto & \frac{1}{\sqrt{\operatorname{det}(\vecC_S(\vecnu))^N
\operatorname{det}(\vecC^m_S(\vecnu))}}
\\
&&{}\times\exp\biggl[-\frac{1}{2}\bigl(\widetilde{\vecX}^f-
\vecmu\bigr)^T\bigl(\vecC _S(\vecnu) \otimes
\vecC_T(\kappa)\bigr)^{-1}\bigl(\widetilde{
\vecX}^f-\vecmu \bigr)
\\
&&\hspace*{11pt}\qquad {}-\frac{1}{2}\bigl(\vecx^m-\vecxi\vecomega
\bigr)^T\vecC^m_S(\vecnu )^{-1}
\bigl(\vecx^m-\vecxi\vecomega\bigr)\biggr]p(\nu_i).
\end{eqnarray*}

Here $|\cdot$ denotes conditioning on the rest of the parameters and
the data.
Note in the above formulas that $\veclambda^f_{ci}$ depends on the past
temperatures
$\widetilde{x}^f_{ci}$
and both $\veclambda^m_{i}$ and $\veclambda^f_{ci}$ depend on the
temperature response parameters $\alpha_j$, $\beta_j$, $\gamma_j$
(cf. Sections~\ref{multibummer} and \ref{spat-tempmodel}). In MCMC
simulation, the probabilities $\vecp_i^m$ and $\vecp_{ci}^f$ as well as
the spatial trend parameter $\vecomega$ can be updated using Gibbs
sampling while all other parameters are updated using the
Metropolis--Hastings algorithm.
\end{appendix}

\section*{Acknowledgments}
We are grateful to Dr. Caspar Ammann from NCAR who provided us with
the simulated temperature times series used in Section~\ref{EstimationOfKappa}.

\begin{supplement}[id=suppA]
\sname{Supplement A}
\stitle{Additional analyses, reconstructions and description of the data}
\slink[doi]{10.1214/15-AOAS832SUPPA} %[doi,text={...}] - jei reikia
%suskaldyti doi
\sdatatype{.pdf}
\sfilename{aoas832\_suppa.pdf}
\sdescription{The document (a pdf-file) includes an analysis of the
Gaussian response model and its parameters, reference records from
Greenland ice cores and Scandinavian lake sediments, additional
reconstructions, a list of the core chronologies for the four lakes
used for temperature reconstruction, and charts of relative abundances
of the ten most common pollen taxa in the samples.}
\end{supplement}

\begin{supplement}[id=suppB]
\sname{Supplement B}
\stitle{The data}
\slink[doi]{10.1214/15-AOAS832SUPPB} %[doi,text={...}] - jei reikia
%suskaldyti doi
\sdatatype{.zip}
\sfilename{aoas832\_suppb.zip}
\sdescription{The data used in the article (an Excel file).}
\end{supplement}

\begin{supplement}[id=suppC]
\sname{Supplement C}
\stitle{The Matlab code}
\slink[doi]{10.1214/15-AOAS832SUPPC} %[doi,text={...}] - jei reikia
%suskaldyti doi
\sdatatype{.zip}
\sfilename{aoas832\_suppc.zip}
\sdescription{The Matlab code used in reconstructions.}
\end{supplement}

% imsref loaded by daiva.urboniene, 2015-06-17 12:48:06

%
%\section{}
%\end{appendix}

%\section*{Acknowledgments}

%\begin{thebibliography}{99}
%\bibitem{r1}
%\bibitem{r1}
%\end{thebibliography}

\printaddresses

\begin{thebibliography}{47}
% pybtex-1.35. Style name=ims, version=2.92, label_style=nameyear, sorting_style=complex, cfg=None, language=None.


%b1 ###
\bibitem[\protect\citeauthoryear{Alley and Ag{\'u}stsd{\'o}ttir}{2005}]{AllAgu05}
\begin{barticle}[author]
\bauthor{\bsnm{Alley},~\bfnm{R.~B.}\binits{R.~B.}} \AND
\bauthor{\bsnm{Ag{\'u}stsd{\'o}ttir},~\bfnm{A.~M.}\binits{A.~M.}}
(\byear{2005}).
\btitle{The 8k event: Cause and consequences of a~major Holocene abrupt climate change}.
\bjournal{Quat. Sci. Rev.}
\bvolume{24}
\bpages{1123--1149}.
\end{barticle}
%

\bptok{imsref}%
\endbibitem

%b2 ###
\bibitem[\protect\citeauthoryear{Ammann et~al.}{2007}]{Ammetal07}
\begin{barticle}[author]
\bauthor{\bsnm{Ammann},~\bfnm{C.~M.}\binits{C.~M.}},
\bauthor{\bsnm{Joos},~\bfnm{F.}\binits{F.}},
\bauthor{\bsnm{Schimel},~\bfnm{D.~S.}\binits{D.~S.}},
\bauthor{\bsnm{Otto-Bliesner},~\bfnm{B.~L.}\binits{B.~L.}} \AND
\bauthor{\bsnm{Tomas},~\bfnm{R.~A.}\binits{R.~A.}}
(\byear{2007}).
\btitle{Solar influence on climate during the past millennium: Results from transient simulations with the NCAR climate system model}.
\bjournal{Proc. Natl. Acad. Sci. USA}
\bvolume{104}
\bpages{3713--3718}.
\end{barticle}
%

\bptok{imsref}%
\endbibitem

%b3 ###
\bibitem[\protect\citeauthoryear{Antonsson et~al.}{2006}]{AntonssonEtAl2006}
\begin{barticle}[author]
\bauthor{\bsnm{Antonsson},~\bfnm{K.}\binits{K.}},
\bauthor{\bsnm{Brooks},~\bfnm{S.~J.}\binits{S.~J.}},
\bauthor{\bsnm{Sepp{\"{a}}},~\bfnm{H.}\binits{H.}},
\bauthor{\bsnm{Telford},~\bfnm{R.~J.}\binits{R.~J.}} \AND
\bauthor{\bsnm{Birks},~\bfnm{H.~J.~B.}\binits{H.~J.~B.}}
(\byear{2006}).
\btitle{Quantitative palaeotemperature records iferred from fossil chironomid and pollen assemblages from {Lake Gilltj\"arnen}, northern central {Sweden}}.
\bjournal{J. Quat. Sci.}
\bvolume{21}
\bpages{831--841}.
\end{barticle}
%

\bptok{imsref}%
\endbibitem

%b4 ###
\bibitem[\protect\citeauthoryear{Banerjee, Carlin and Gelfand}{2004}]{BanerjeeEtAl}
\begin{bbook}[author]
\bauthor{\bsnm{Banerjee},~\bfnm{S.}\binits{S.}},
\bauthor{\bsnm{Carlin},~\bfnm{B.~P.}\binits{B.~P.}} \AND
\bauthor{\bsnm{Gelfand},~\bfnm{A.~E.}\binits{A.~E.}}
(\byear{2004}).
\btitle{Hierarchical Modeling and Analysis for Spatial Data}.
\bpublisher{Chapman {\&} Hall},
\blocation{London}.
\end{bbook}
%

\bptok{imsref}%
\endbibitem

%b5 ###
\bibitem[\protect\citeauthoryear{Birks et~al.}{2010}]{BirksEtAl2010}
\begin{barticle}[author]
\bauthor{\bsnm{Birks},~\bfnm{H.~J.~B.}\binits{H.~J.~B.}},
\bauthor{\bsnm{Heiri},~\bfnm{O.}\binits{O.}},
\bauthor{\bsnm{Sepp{\"{a}}},~\bfnm{H.}\binits{H.}} \AND
\bauthor{\bsnm{Bjune},~\bfnm{A.~E.}\binits{A.~E.}}
(\byear{2010}).
\btitle{Strengths and weaknesses of quantitative climate reconstructions based on late-quaternary biological proxies}.
\bjournal{The Open Ecology Jounal}
\bvolume{3}
\bpages{68--110}.
\end{barticle}
%

\bptok{imsref}%
\endbibitem

%b6 ###
\bibitem[\protect\citeauthoryear{Brynjarsd{\'o}ttir and Berliner}{2011}]{BrynjarsdottirBerliner2011}
\begin{barticle}[mr]
\bauthor{\bsnm{Brynjarsd{\'o}ttir},~\bfnm{Jenn{\'y}}\binits{J.}} \AND
\bauthor{\bsnm{Berliner},~\bfnm{L.~Mark}\binits{L.~M.}}
(\byear{2011}).
\btitle{Bayesian hierarchical modeling for temperature reconstruction from geothermal data}.
\bjournal{Ann. Appl. Stat.}
\bvolume{5}
\bpages{1328--1359}.
\bid{doi={10.1214/10-AOAS452}, issn={1932-6157}, mr={2849776}}
\end{barticle}
%

\bptok{imsref}%
% NOT OUTPUTTED:
%   number = 2B
%   doi = http://dx.doi.org/10.1214/10-AOAS452
%   fjournal = The Annals of Applied Statistics
\endbibitem

%b7 ###
\bibitem[\protect\citeauthoryear{Cressie}{1993}]{Cressie}
\begin{bbook}[mr]
\bauthor{\bsnm{Cressie},~\bfnm{Noel~A.~C.}\binits{N.~A.~C.}}
(\byear{1993}).
\btitle{Statistics for Spatial Data}.
%\bseries{Wiley Series in Probability and Mathematical Statistics: Applied Probability and Statistics}.
\bpublisher{Wiley},
\blocation{New York}.
%\bnote{Revised reprint of the 1991 edition, A Wiley-Interscience Publication}.
\bid{doi={10.1002/9781119115151}, mr={1239641}}
\end{bbook}
%

\bptok{imsref}%
% NOT OUTPUTTED:
%   doi = http://dx.doi.org/10.1002/9781119115151
%   isbn = 0-471-00255-0
%   fpage = xxii+900
\endbibitem

%b8 ###
\bibitem[\protect\citeauthoryear{Dahl}{1998}]{Dahl1998}
\begin{bbook}[author]
\bauthor{\bsnm{Dahl},~\bfnm{E.}\binits{E.}}
(\byear{1998}).
\btitle{The Phytogeography of Northern Europe: British Isles, Fennoscandia, and Adjacent Areas}.
\bpublisher{Cambridge Univ. Press},
\blocation{Cambridge}.
\end{bbook}
%

\bptok{imsref}%
\endbibitem

%b9 ###
\bibitem[\protect\citeauthoryear{Er{\"a}st{\"o} and Holmstr{\"o}m}{2005}]{ErastoHolmstrom2005}
\begin{barticle}[mr]
\bauthor{\bsnm{Er{\"a}st{\"o}},~\bfnm{Panu}\binits{P.}} \AND
\bauthor{\bsnm{Holmstr{\"o}m},~\bfnm{Lasse}\binits{L.}}
(\byear{2005}).
\btitle{Bayesian multiscale smoothing for making inferences about features in scatterplots}.
\bjournal{J. Comput. Graph. Statist.}
\bvolume{14}
\bpages{569--589}.
\bid{doi={10.1198/106186005X59315}, issn={1061-8600}, mr={2170202}}
\bptnote{check pages}%
\end{barticle}
%

\bptok{imsref}%
% NOT OUTPUTTED:
%   number = 3
%   doi = http://dx.doi.org/10.1198/106186005X59315
%   fjournal = Journal of Computational and Graphical Statistics
\endbibitem

%b10 ###
\bibitem[\protect\citeauthoryear{Er{\"{a}}st{\"{o}} and Holmstr{\"{o}}m}{2006}]{ErastoHolmstrom2006}
\begin{barticle}[author]
\bauthor{\bsnm{Er{\"{a}}st{\"{o}}},~\bfnm{P.}\binits{P.}} \AND
\bauthor{\bsnm{Holmstr{\"{o}}m},~\bfnm{L.}\binits{L.}}
(\byear{2006}).
\btitle{Selection of prior distributions and multiscale analysis in {B}ayesian temperature reconstructions based on fossil assemblages}.
\bjournal{J. Paleolimnol.}
\bvolume{36}
\bpages{69--80}.
\end{barticle}
%

\bptok{imsref}%
\endbibitem

%b11 ###
\bibitem[\protect\citeauthoryear{Er{\"a}st{\"o} et~al.}{2012}]{ErastoEtAl2012}
\begin{barticle}[mr]
\bauthor{\bsnm{Er{\"a}st{\"o}},~\bfnm{Panu}\binits{P.}},
\bauthor{\bsnm{Holmstr{\"o}m},~\bfnm{Lasse}\binits{L.}},
\bauthor{\bsnm{Korhola},~\bfnm{Atte}\binits{A.}} \AND
\bauthor{\bsnm{Weckstr{\"o}m},~\bfnm{Jan}\binits{J.}}
(\byear{2012}).
\btitle{Finding a consensus on credible features among several paleoclimate reconstructions}.
\bjournal{Ann. Appl. Stat.}
\bvolume{6}
\bpages{1377--1405}.
\bid{doi={10.1214/12-AOAS540}, issn={1932-6157}, mr={3058668}}\vadjust{\goodbreak}
\end{barticle}
%

\bptok{imsref}%
% NOT OUTPUTTED:
%   number = 4
%   doi = http://dx.doi.org/10.1214/12-AOAS540
%   fjournal = The Annals of Applied Statistics
\endbibitem

%b12 ###
\bibitem[\protect\citeauthoryear{Gelman et~al.}{2004}]{Gelman}
\begin{bbook}[mr]
\bauthor{\bsnm{Gelman},~\bfnm{Andrew}\binits{A.}},
\bauthor{\bsnm{Carlin},~\bfnm{John~B.}\binits{J.~B.}},
\bauthor{\bsnm{Stern},~\bfnm{Hal~S.}\binits{H.~S.}} \AND
\bauthor{\bsnm{Rubin},~\bfnm{Donald~B.}\binits{D.~B.}}
(\byear{2004}).
\btitle{Bayesian Data Analysis},
\bedition{2nd} ed.
%\bseries{Texts in Statistical Science Series}.
\bpublisher{Chapman \& Hall/CRC},
\blocation{Boca Raton, FL}.
\bid{mr={2027492}}
\end{bbook}
%

\bptok{imsref}%
% NOT OUTPUTTED:
%   isbn = 1-58488-388-X
%   fpage = xxvi+668
\endbibitem

%b13 ###
\bibitem[\protect\citeauthoryear{Haslett and Parnell}{2008}]{HasPar08}
\begin{barticle}[mr]
\bauthor{\bsnm{Haslett},~\bfnm{John}\binits{J.}} \AND
\bauthor{\bsnm{Parnell},~\bfnm{Andrew}\binits{A.}}
(\byear{2008}).
\btitle{A simple monotone process with application to radiocarbon-dated depth chronologies}.
\bjournal{J. Roy. Statist. Soc. Ser. C}
\bvolume{57}
\bpages{399--418}.
\bid{doi={10.1111/j.1467-9876.2008.00623.x}, issn={0035-9254}, mr={2526125}}
\bptnote{check pages}%
\end{barticle}
%

\bptok{imsref}%
% NOT OUTPUTTED:
%   number = 4
%   doi = http://dx.doi.org/10.1111/j.1467-9876.2008.00623.x
%   fjournal = Journal of the Royal Statistical Society. Series C. Applied Statistics
\endbibitem

%b14 ###
\bibitem[\protect\citeauthoryear{Haslett et~al.}{2006}]{HaslettEtAl2006}
\begin{barticle}[mr]
\bauthor{\bsnm{Haslett},~\bfnm{J.}\binits{J.}},
\bauthor{\bsnm{Whiley},~\bfnm{M.}\binits{M.}},
\bauthor{\bsnm{Bhattacharya},~\bfnm{S.}\binits{S.}},
\bauthor{\bsnm{Salter-Townshend},~\bfnm{M.}\binits{M.}},
\bauthor{\bsnm{Wilson},~\bfnm{Simon~P.}\binits{S.~P.}},
\bauthor{\bsnm{Allen},~\bfnm{J.~R.~M.}\binits{J.~R.~M.}},
\bauthor{\bsnm{Huntley},~\bfnm{B.}\binits{B.}} \AND
\bauthor{\bsnm{Mitchell},~\bfnm{F.~J.~G.}\binits{F.~J.~G.}}
(\byear{2006}).
\btitle{Bayesian palaeoclimate reconstruction}.
\bjournal{J. Roy. Statist. Soc. Ser. A}
\bvolume{169}
\bpages{395--438}.
\bid{doi={10.1111/j.1467-985X.2006.00429.x}, issn={0964-1998}, mr={2236914}}
\end{barticle}
%

\bptok{imsref}%
% NOT OUTPUTTED:
%   number = 3
%   doi = http://dx.doi.org/10.1111/j.1467-985X.2006.00429.x
%   fjournal = Journal of the Royal Statistical Society. Series A. Statistics in Society
\endbibitem



\bibitem[\protect\citeauthoryear{Holmstr{\"o}m et~al.}{2015a}]{Holetal14suppa}
\begin{bmisc}[author]
{\bauthor{\bsnm{Holmstr\"om},~\binits{L.}},
\bauthor{\bsnm{Ilvonen},~\binits{L.}},
\bauthor{\bsnm{Sepp\"a,~\binits{H.}}} \AND
\bauthor{\bsnm{Veski},~\binits{S.}}}
(\byear{2015a}).
\bhowpublished{Supplement A to ``A Bayesian spatiotemporal model for reconstructing climate from
multiple pollen records.''
DOI:\doiurl{10.1214/15-AOAS832SUPPA}}.
\bnote{An on line supplement}.
\bptok{imsref}%
\end{bmisc}

\bptok{imsref}%
\endbibitem

%b16 ###
\bibitem[\protect\citeauthoryear{Holmstr{\"o}m et~al.}{2015b}]{Holetal14suppb}
\begin{bmisc}[author]
{\bauthor{\bsnm{Holmstr\"om},~\binits{L.}},
\bauthor{\bsnm{Ilvonen},~\binits{L.}},
\bauthor{\bsnm{Sepp\"a,~\binits{H.}}} \AND
\bauthor{\bsnm{Veski},~\binits{S.}}}
(\byear{2015b}).
\bhowpublished{Supplement B to ``A Bayesian spatiotemporal model for reconstructing climate from
multiple pollen records.''
DOI:\doiurl{10.1214/15-AOAS832SUPPB}}.
\bnote{The data used in reconstructions}.
\end{bmisc}
%

\bptok{imsref}%
\endbibitem

%b17 ###
\bibitem[\protect\citeauthoryear{Holmstr{\"o}m et~al.}{2015c}]{Holetal14suppc}
\begin{bmisc}[author]
{\bauthor{\bsnm{Holmstr\"om},~\binits{L.}},
\bauthor{\bsnm{Ilvonen},~\binits{L.}},
\bauthor{\bsnm{Sepp\"a,~\binits{H.}}} \AND
\bauthor{\bsnm{Veski},~\binits{S.}}}
(\byear{2015c}).
\bhowpublished{Supplement C to ``A Bayesian spatiotemporal model for reconstructing climate from
multiple pollen records.''
DOI:\doiurl{10.1214/15-AOAS832SUPPC}}.
\bnote{The Matlab code used in reconstructions}.
\end{bmisc}
%

\bptok{imsref}%
\endbibitem

%b18 ###
\bibitem[\protect\citeauthoryear{Jansen et~al.}{2007}]{Janetal2007}
\begin{bincollection}[author]
\bauthor{\bsnm{Jansen},~\bfnm{E.}\binits{E.}} \betal{et~al.}
(\byear{2007}).
\btitle{Palaeoclimate}.
In \bbooktitle{Climate Change 2007: The Physical Science Basis. Contribution of Working Group I
to the Fourth Assessment Report of the Intergovernmental Panel on Climate Change}
(\beditor{\bfnm{S.}\binits{S.}~\bsnm{Solomon}},
\beditor{\bfnm{D.}\binits{D.}~\bsnm{Qin}},
\beditor{\bfnm{M.}\binits{M.}~\bsnm{Manning}},
\beditor{\bfnm{Z.}\binits{Z.}~\bsnm{Chen}},
\beditor{\bfnm{M.}\binits{M.}~\bsnm{Marquis}},
\beditor{\bfnm{K.~B.}\binits{K.~B.}~\bsnm{Averyt}},
\beditor{\bfnm{M.}\binits{M.}~\bsnm{Tignor}} \AND
\beditor{\bfnm{H.~L.}\binits{H.~L.}~\bsnm{Miller}}, eds.)
\bpages{433--497}.
\bpublisher{Cambridge Univ. Press},
\blocation{Cambridge}.
\end{bincollection}
%

\bptok{imsref}%
\endbibitem

%b19 ###
\bibitem[\protect\citeauthoryear{Jones et~al.}{2009}]{JonesEtAl2009}
\begin{barticle}[author]
\bauthor{\bsnm{Jones},~\bfnm{P.~D.}\binits{P.~D.}},
\bauthor{\bsnm{Briffa},~\bfnm{P.~D.}\binits{P.~D.}},
\bauthor{\bsnm{Osborn},~\bfnm{T.~J.}\binits{T.~J.}},
\bauthor{\bsnm{Lough},~\bfnm{J.~M.}\binits{J.~M.}},
\bauthor{\bparticle{van} \bsnm{Ommen},~\bfnm{T.~D.}\binits{T.~D.}},
\bauthor{\bsnm{Vinther},~\bfnm{B.~M.}\binits{B.~M.}},
\bauthor{\bsnm{Luterbacher},~\bfnm{J.}\binits{J.}},
\bauthor{\bsnm{Wahl},~\bfnm{E.~R.}\binits{E.~R.}},
\bauthor{\bsnm{Zwiers},~\bfnm{F.~W.}\binits{F.~W.}},
\bauthor{\bsnm{Mann},~\bfnm{M.~E.}\binits{M.~E.}},
\bauthor{\bsnm{Schmidt},~\bfnm{G.~A.}\binits{G.~A.}},
\bauthor{\bsnm{Ammann},~\bfnm{C.~M.}\binits{C.~M.}},
\bauthor{\bsnm{Buckley},~\bfnm{B.~M.}\binits{B.~M.}},
\bauthor{\bsnm{Cobb},~\bfnm{K.~M.}\binits{K.~M.}},
\bauthor{\bsnm{Esper},~\bfnm{J.}\binits{J.}},
\bauthor{\bsnm{Goosse},~\bfnm{H.}\binits{H.}},
\bauthor{\bsnm{Graham},~\bfnm{N.}\binits{N.}},
\bauthor{\bsnm{Jansen},~\bfnm{E.}\binits{E.}},
\bauthor{\bsnm{Kiefer},~\bfnm{T.}\binits{T.}},
\bauthor{\bsnm{Kull},~\bfnm{C.}\binits{C.}},
\bauthor{\bsnm{K{\"{u}}ttel},~\bfnm{M.}\binits{M.}},
\bauthor{\bsnm{Mosley-Thompson},~\bfnm{E.}\binits{E.}},
\bauthor{\bsnm{Overpeck},~\bfnm{J.~T.}\binits{J.~T.}},
\bauthor{\bsnm{Riedwyl},~\bfnm{N.}\binits{N.}},
\bauthor{\bsnm{Schulz},~\bfnm{M.}\binits{M.}},
\bauthor{\bsnm{Tudhope},~\bfnm{A.~W.}\binits{A.~W.}},
\bauthor{\bsnm{Villalba},~\bfnm{R.}\binits{R.}},
\bauthor{\bsnm{Wanner},~\bfnm{H.}\binits{H.}},
\bauthor{\bsnm{Wolff},~\bfnm{E.}\binits{E.}} \AND
\bauthor{\bsnm{Xoplaki},~\bfnm{E.}\binits{E.}}
(\byear{2009}).
\btitle{High-resolution palaeoclimatology of the last millennium: A review of current status and future prospects}.
\bjournal{Holocene}
\bvolume{19}
\bpages{3--49}.
\end{barticle}
%

\bptok{imsref}%
\endbibitem

%b20 ###
\bibitem[\protect\citeauthoryear{Journel and Huijbregts}{1978}]{Journel1978}
\begin{bbook}[author]
\bauthor{\bsnm{Journel},~\bfnm{A.~G.}\binits{A.~G.}} \AND
\bauthor{\bsnm{Huijbregts},~\bfnm{C.~J.}\binits{C.~J.}}
(\byear{1978}).
\btitle{Mining Geostatistics}.
\bpublisher{Academic Press},
\blocation{San Diego}.
\end{bbook}
%

\bptok{imsref}%
\endbibitem

%b21 ###
\bibitem[\protect\citeauthoryear{Juggins and Birks}{2012}]{JugBir2012}
\begin{bincollection}[author]
\bauthor{\bsnm{Juggins},~\bfnm{S.}\binits{S.}} \AND
\bauthor{\bsnm{Birks},~\bfnm{H.~J.~B.}\binits{H.~J.~B.}}
(\byear{2012}).
\btitle{Quantitative environmental reconstructions from biological data}.
In \bbooktitle{Tracking Environmental Change Using Lake Sediments},
\bseries{Data Handling and Numerical Techniques}
\bvolume{5}
(\beditor{\bfnm{H.~J.~B.}\binits{H.~J.~B.}~\bsnm{Birks}},
\beditor{\bfnm{A.~F.}\binits{A.~F.}~\bsnm{Lotter}},
\beditor{\bfnm{S.}\binits{S.}~\bsnm{Juggins}} \AND
\beditor{\bfnm{J.~P.}\binits{J.~P.}~\bsnm{Smol}}, eds.)
\bpages{431--494}.
\bpublisher{Springer},
\blocation{Dordrecht}.
\end{bincollection}
%

\bptok{imsref}%
\endbibitem

%b22 ###
\bibitem[\protect\citeauthoryear{Kaipio and Somersalo}{2005}]{KaipioSomersalo}
\begin{bbook}[mr]
\bauthor{\bsnm{Kaipio},~\bfnm{Jari}\binits{J.}} \AND
\bauthor{\bsnm{Somersalo},~\bfnm{Erkki}\binits{E.}}
(\byear{2005}).
\btitle{Statistical and Computational Inverse Problems}.
\bseries{Applied Mathematical Sciences}
\bvolume{160}.
\bpublisher{Springer},
\blocation{New York}.
\bid{mr={2102218}}
\end{bbook}
%

\bptok{imsref}%
% NOT OUTPUTTED:
%   isbn = 0-387-22073-9
%   fpage = xvi+339
\endbibitem

%b23 ###
\bibitem[\protect\citeauthoryear{Korhola et~al.}{2002}]{KorholaEtAl2002}
\begin{barticle}[author]
\bauthor{\bsnm{Korhola},~\bfnm{A.}\binits{A.}},
\bauthor{\bsnm{Vasko},~\bfnm{K.}\binits{K.}},
\bauthor{\bsnm{Toivonen},~\bfnm{H.~T.~T.}\binits{H.~T.~T.}} \AND
\bauthor{\bsnm{Olander},~\bfnm{H.}\binits{H.}}
(\byear{2002}).
\btitle{Holocene temperature changes in northern {F}ennoscandia reconstructed from chironomids using {B}ayesian modelling}.
\bjournal{Qaternary Science Reviews}
\bvolume{21}
\bpages{1841--1860}.
\end{barticle}
%

\bptok{imsref}%
\endbibitem

%b24 ###
\bibitem[\protect\citeauthoryear{Li, Nychka and Ammann}{2010}]{LiEtAl2010}
\begin{barticle}[author]
\bauthor{\bsnm{Li},~\bfnm{B.}\binits{B.}},
\bauthor{\bsnm{Nychka},~\bfnm{D.~W.}\binits{D.~W.}} \AND
\bauthor{\bsnm{Ammann},~\bfnm{C.~M.}\binits{C.~M.}}
(\byear{2010}).
\btitle{The value of multi-proxy reconstruction of past climate}.
\bjournal{J. Amer. Statist. Assoc.}
\bvolume{105}
\bpages{883--911}.
\end{barticle}
%

\bptok{imsref}%
\endbibitem

%b25 ###
\bibitem[\protect\citeauthoryear{Marcott et~al.}{2013}]{Mar2013}
\begin{barticle}[pbm]
\bauthor{\bsnm{Marcott},~\bfnm{Shaun~A.}\binits{S.~A.}},
\bauthor{\bsnm{Shakun},~\bfnm{Jeremy~D.}\binits{J.~D.}},
\bauthor{\bsnm{Clark},~\bfnm{Peter~U.}\binits{P.~U.}} \AND
\bauthor{\bsnm{Mix},~\bfnm{Alan~C.}\binits{A.~C.}}
(\byear{2013}).
\btitle{A reconstruction of regional and global temperature for the past 11{,}300 years}.
\bjournal{Science}
\bvolume{339}
\bpages{1198--1201}.
\bid{doi={10.1126/science.1228026}, issn={1095-9203}, pii={339/6124/1198}, pmid={23471405}}
\end{barticle}
%

\bptok{imsref}%
% NOT OUTPUTTED:
%   number = 6124
%   fjournal = Science (New York, N.Y.)
\endbibitem

%b26 ###
\bibitem[\protect\citeauthoryear{Masson-Delmotte et~al.}{2013}]{Masetal2013}
\begin{bincollection}[author]
\bauthor{\bsnm{Masson-Delmotte},~\bfnm{V.}\binits{V.}} \betal{et~al.}
(\byear{2013}).
\btitle{Information from paleoclimate archives}.
In \bbooktitle{Climate Change 2013: The Physical Science Basis. Contribution of Working Group I to the Fifth Assessment Report of the Intergovernmental Panel on Climate Change}
(\beditor{\bfnm{T.~F.}\binits{T.~F.}~\bsnm{Stocker}},
\beditor{\bfnm{G.~K.}\binits{G.~K.}~\bsnm{Plattner}},
\beditor{\bfnm{M.}\binits{M.}~\bsnm{Tignor}},
\beditor{\bfnm{S.~K.}\binits{S.~K.}~\bsnm{Allen}},
\beditor{\bfnm{J.}\binits{J.}~\bsnm{Boschung}},
\beditor{\bfnm{A.}\binits{A.}~\bsnm{Nauels}},
\beditor{\bfnm{Y.}\binits{Y.}~\bsnm{Xia}} \AND
\beditor{\bfnm{P.~M.}\binits{P.~M.}~\bsnm{Midgley}}, eds.)
\bpages{383--464}.
\bpublisher{Cambridge Univ. Press},
\blocation{Cambridge}.
\end{bincollection}
%

\bptok{imsref}%
\endbibitem

%b27 ###
\bibitem[\protect\citeauthoryear{Moberg and Bergstr{\"o}m}{1997}]{MobergBergstrom1997}
\begin{barticle}[author]
\bauthor{\bsnm{Moberg},~\bfnm{A.}\binits{A.}} \AND
\bauthor{\bsnm{Bergstr{\"o}m},~\bfnm{H.}\binits{H.}}
(\byear{1997}).
\btitle{Homogenization of Swedish temperature data. Part III: The long temperature records from Uppsala and {Stockholm}}.
\bjournal{Int. J. Climatol.}
\bvolume{17}\vadjust{\goodbreak}
\bpages{667--699}.
\end{barticle}
%

\bptok{imsref}%
\endbibitem

%b28 ###
\bibitem[\protect\citeauthoryear{NRC}{2006}]{NRC}
\begin{bmisc}[author]
\borganization{NRC}
(\byear{2006}).
\bhowpublished{\textit{Surface Temperature Reconstructions for the Last 2000 Years}.
The National Academies Press,
Washington.}
\end{bmisc}
%

\bptok{imsref}%
\endbibitem

%b29 ###
\bibitem[\protect\citeauthoryear{Ohlwein and Wahl}{2012}]{OhlWah2012}
\begin{barticle}[author]
\bauthor{\bsnm{Ohlwein},~\bfnm{C.}\binits{C.}} \AND
\bauthor{\bsnm{Wahl},~\bfnm{E.~R.}\binits{E.~R.}}
(\byear{2012}).
\btitle{Review of probabilistic pollen-climate transfer methods}.
\bjournal{Quat. Sci. Rev.}
\bvolume{31}
\bpages{17--29}.
\bid{doi={10.1016/j.quascirev.2011.11.002}}
\end{barticle}
%

\bptok{imsref}%
\endbibitem

%b30 ###
\bibitem[\protect\citeauthoryear{Paciorek and McLachlan}{2009}]{Pac2009}
\begin{barticle}[mr]
\bauthor{\bsnm{Paciorek},~\bfnm{Christopher~J.}\binits{C.~J.}} \AND
\bauthor{\bsnm{McLachlan},~\bfnm{Jason~S.}\binits{J.~S.}}
(\byear{2009}).
\btitle{Mapping ancient forests: {B}ayesian inference for spatio-temporal trends in forest composition using the fossil pollen proxy record}.
\bjournal{J. Amer. Statist. Assoc.}
\bvolume{104}
\bpages{608--622}.
\bid{doi={10.1198/jasa.2009.0026}, issn={0162-1459}, mr={2751442}}
\end{barticle}
%

\bptok{imsref}%
% NOT OUTPUTTED:
%   number = 486
%   doi = http://dx.doi.org/10.1198/jasa.2009.0026
%   coden = JSTNAL
%   fjournal = Journal of the American Statistical Association
\endbibitem

%b32 ###
\bibitem[\protect\citeauthoryear{Renssen et~al.}{2012}]{Renetal12}
\begin{barticle}[author]
\bauthor{\bsnm{Renssen},~\bfnm{H.}\binits{H.}},
\bauthor{\bsnm{Sepp{\"a}},~\bfnm{H.}\binits{H.}},
\bauthor{\bsnm{Crosta},~\bfnm{X.}\binits{X.}},
\bauthor{\bsnm{Goosse},~\bfnm{H.}\binits{H.}} \AND
\bauthor{\bsnm{Roche},~\bfnm{D.~M.}\binits{D.~M.}}
(\byear{2012}).
\btitle{Global characterization of the {H}olocene thermal maximum}.
\bjournal{Quat. Sci. Rev.}
\bvolume{48}
\bpages{7--19}.
\end{barticle}
%

\bptok{imsref}%
\endbibitem


%b31 ###
\bibitem[\protect\citeauthoryear{Renssen et~al.}{2009}]{Renetal09}
\begin{barticle}[author]
\bauthor{\bsnm{Renssen},~\bfnm{H.}\binits{H.}},
\bauthor{\bsnm{Sepp{\"a}},~\bfnm{H.}\binits{H.}},
\bauthor{\bsnm{Heiri},~\bfnm{O.}\binits{O.}},
\bauthor{\bsnm{Roche},~\bfnm{D.~M.}\binits{D.~M.}},
\bauthor{\bsnm{Goosse},~\bfnm{H.}\binits{H.}} \AND
\bauthor{\bsnm{Fichefet},~\bfnm{T.}\binits{T.}}
(\byear{2009}).
\btitle{{The spatial and temporal complexity of the Holocene thermal maximum}}.
\bjournal{Nat. Geosci.}
\bvolume{2}
\bpages{411--414}.
\end{barticle}
%

\bptok{imsref}%
\endbibitem



%b33 ###
\bibitem[\protect\citeauthoryear{Robert and Casella}{2004}]{RobertCasella}
\begin{bbook}[mr]
\bauthor{\bsnm{Robert},~\bfnm{Christian~P.}\binits{C.~P.}} \AND
\bauthor{\bsnm{Casella},~\bfnm{George}\binits{G.}}
(\byear{2004}).
\btitle{Monte {C}arlo Statistical Methods},
\bedition{2nd} ed.
%\bseries{Springer Texts in Statistics}.
\bpublisher{Springer},
\blocation{New York}.
\bid{doi={10.1007/978-1-4757-4145-2}, mr={2080278}}
\end{bbook}
%

\bptok{imsref}%
% NOT OUTPUTTED:
%   doi = http://dx.doi.org/10.1007/978-1-4757-4145-2
%   isbn = 0-387-21239-6
%   fpage = xxx+645
\endbibitem

%b34 ###
\bibitem[\protect\citeauthoryear{Salonen et~al.}{2012}]{SalonenEtAl2012}
\begin{barticle}[author]
\bauthor{\bsnm{Salonen},~\bfnm{J.~S.}\binits{J.~S.}},
\bauthor{\bsnm{Ilvonen},~\bfnm{L.}\binits{L.}},
\bauthor{\bsnm{Sepp{\"{a}}},~\bfnm{H.}\binits{H.}},
\bauthor{\bsnm{Holmstr{\"{o}}m},~\bfnm{L.}\binits{L.}},
\bauthor{\bsnm{Telford},~\bfnm{R.~J.}\binits{R.~J.}},
\bauthor{\bsnm{Gaidamavi{\v{c}}ius},~\bfnm{A.}\binits{A.}},
\bauthor{\bsnm{Stan{\v{c}}ikait{\.e}},~\bfnm{M.}\binits{M.}} \AND
\bauthor{\bsnm{Subetto},~\bfnm{D.}\binits{D.}}
(\byear{2012}).
\btitle{Comparing different calibration methods ({WA/WA-PLS} regression and {B}ayesian modelling) and
different-sized calibration sets in pollen-based quantitative cllimate reconstructions}.
\bjournal{Holocene}
\bvolume{22}
\bpages{413--424}.
\end{barticle}
%

\bptok{imsref}%
\endbibitem

%b35 ###
\bibitem[\protect\citeauthoryear{Sarmaja-Korjonen and Sepp{\"{a}}}{2007}]{SarmajaKorjonenSeppa2007}
\begin{barticle}[author]
\bauthor{\bsnm{Sarmaja-Korjonen},~\bfnm{K.}\binits{K.}} \AND
\bauthor{\bsnm{Sepp{\"{a}}},~\bfnm{H.}\binits{H.}}
(\byear{2007}).
\btitle{Abrupt and consistent responses of aquatic and terrestrial ecosystems to the 8200 cal. yr BP cold event: A lacustrine record from {Lake Arapisto, Finland}}.
\bjournal{Holocene}
\bvolume{17}
\bpages{455--464}.
\end{barticle}
%

\bptok{imsref}%
\endbibitem

%b36 ###
\bibitem[\protect\citeauthoryear{Sepp{\"{a}}, Hammarlund and Antonsson}{2005}]{SeppaEtAl2005}
\begin{barticle}[author]
\bauthor{\bsnm{Sepp{\"{a}}},~\bfnm{H.}\binits{H.}},
\bauthor{\bsnm{Hammarlund},~\bfnm{D.}\binits{D.}} \AND
\bauthor{\bsnm{Antonsson},~\bfnm{K.}\binits{K.}}
(\byear{2005}).
\btitle{Low-frequency and high-frequency changes in temperature and effective humidity during the Holocene in south-central Sweden:
Implications for atmospheric and oceanic forcings of climate}.
\bjournal{Clim. Dyn.}
\bvolume{25}
\bpages{285--297}.
\end{barticle}
%

\bptok{imsref}%
\endbibitem

%b37 ###
\bibitem[\protect\citeauthoryear{Sepp{\"{a}} and Poska}{2004}]{SeppaPoska2004}
\begin{barticle}[author]
\bauthor{\bsnm{Sepp{\"{a}}},~\bfnm{H.}\binits{H.}} \AND
\bauthor{\bsnm{Poska},~\bfnm{A.}\binits{A.}}
(\byear{2004}).
\btitle{Holocene annual mean temperature changes in Estonia and their relationship
to solar insolation and atmospheric circulation patterns}.
\bjournal{Quat. Res.}
\bvolume{61}
\bpages{22--31}.
\end{barticle}
%

\bptok{imsref}%
\endbibitem

%b38 ###
\bibitem[\protect\citeauthoryear{Sepp{\"{a}} et~al.}{2009}]{SeppaEtAl2009}
\begin{barticle}[author]
\bauthor{\bsnm{Sepp{\"{a}}},~\bfnm{H.}\binits{H.}},
\bauthor{\bsnm{Bjune},~\bfnm{A.~E.}\binits{A.~E.}},
\bauthor{\bsnm{Telford},~\bfnm{R.~J.}\binits{R.~J.}},
\bauthor{\bsnm{Birks},~\bfnm{H.~J.~B.}\binits{H.~J.~B.}} \AND
\bauthor{\bsnm{Veski},~\bfnm{S.}\binits{S.}}
(\byear{2009}).
\btitle{Last nine-thousand years of temperature variability in {Northern Europe}}.
\bjournal{Clim. Past}
\bvolume{5}
\bpages{523--535}.
\end{barticle}
%

\bptok{imsref}%
\endbibitem

%b39 ###
\bibitem[\protect\citeauthoryear{Shakun et~al.}{2012}]{Sha2012}
\begin{barticle}[author]
\bauthor{\bsnm{Shakun},~\bfnm{J.~D.}\binits{J.~D.}} \betal{et~al.}
(\byear{2012}).
\btitle{Global warming preceded by increasing carbon dioxide concentrations during the last deglaciation}.
\bjournal{Nature}
\bvolume{484}
\bpages{49--54}.
\end{barticle}
%

\bptok{imsref}%
\endbibitem

%b40 ###
\bibitem[\protect\citeauthoryear{ter Braak and Juggins}{1993}]{terBraakJuggins1993}
\begin{barticle}[author]
\bauthor{\bsnm{ter Braak},~\bfnm{Cajo~J.~F.}\binits{C.~J.~F.}} \AND
\bauthor{\bsnm{Juggins},~\bfnm{Steve}\binits{S.}}
(\byear{1993}).
\btitle{Weighted averaging partial least squares regression ({WA-PLS}): An improved method for reconstructing environmental variables from species assemblages}.
\bjournal{Hydrobiologia}
\bvolume{269/270}
\bpages{485--502}.
\end{barticle}
%

\bptok{imsref}%
\endbibitem

%b43 ###
\bibitem[\protect\citeauthoryear{Tingley et~al.}{2012}]{TingleyEtAl2012}
\begin{barticle}[author]
\bauthor{\bsnm{Tingley},~\bfnm{M.~P.}\binits{M.~P.}},
\bauthor{\bsnm{Craigmile},~\bfnm{P.~F.}\binits{P.~F.}},
\bauthor{\bsnm{Haran},~\bfnm{M.}\binits{M.}},
\bauthor{\bsnm{Li},~\bfnm{B.}\binits{B.}},
\bauthor{\bsnm{Mannshardt-Shamseldin},~\bfnm{E.}\binits{E.}} \AND
\bauthor{\bsnm{Rajaratnam},~\bfnm{B.}\binits{B.}}
(\byear{2012}).
\btitle{Piecing together the past: Statistical insights into paleoclimatic reconstructions}.
\bjournal{Quat. Sci. Rev.}
\bvolume{35}
\bpages{1--22}.
\end{barticle}
%

\bptok{imsref}%
\endbibitem


%b41 ###
\bibitem[\protect\citeauthoryear{Tingley and Huybers}{2010a}]{TingleyHuybers2010A}
\begin{barticle}[author]
\bauthor{\bsnm{Tingley},~\bfnm{M.~P.}\binits{M.~P.}} \AND
\bauthor{\bsnm{Huybers},~\bfnm{P.}\binits{P.}}
(\byear{2010}a).
\btitle{A {B}ayesian algorithm for reconstructing climate anomalies in space and time. {P}art 1: Development and applications to paleoclimate reconstruction problems}.
\bjournal{J. Climate}
\bvolume{23}
\bpages{2759--2781}.
\end{barticle}
%

\bptok{imsref}%
\endbibitem

%b42 ###
\bibitem[\protect\citeauthoryear{Tingley and Huybers}{2010b}]{TingleyHuybers2010B}
\begin{barticle}[author]
\bauthor{\bsnm{Tingley},~\bfnm{M.~P.}\binits{M.~P.}} \AND
\bauthor{\bsnm{Huybers},~\bfnm{P.}\binits{P.}}
(\byear{2010}b).
\btitle{A {B}ayesian algorithm for reconstructing climate anomalies in space and time. Part 2: Comparison with the regularized expectation-maximization algorithm}.
\bjournal{Journal of Climate}
\bvolume{23}
\bpages{2782--2800}.
\end{barticle}
%

\bptok{imsref}%
\endbibitem



%b44 ###
\bibitem[\protect\citeauthoryear{Toivonen et~al.}{2001}]{ToivonenEtAl2001}
\begin{barticle}[author]
\bauthor{\bsnm{Toivonen},~\bfnm{H.~T.~T.}\binits{H.~T.~T.}},
\bauthor{\bsnm{Mannila},~\bfnm{H.}\binits{H.}},
\bauthor{\bsnm{Korhola},~\bfnm{A.}\binits{A.}} \AND
\bauthor{\bsnm{Olander},~\bfnm{H.}\binits{H.}}
(\byear{2001}).
\btitle{Applying {B}ayesian statistics to organism-based environmental reconstruction}.
\bjournal{Ecol. Appl.}
\bvolume{11}
\bpages{618--630}.
\end{barticle}
%

\bptok{imsref}%
\endbibitem

%b45 ###
\bibitem[\protect\citeauthoryear{Vasko, Toivonen and Korhola}{2000}]{VaskoEtAl2000}
\begin{barticle}[author]
\bauthor{\bsnm{Vasko},~\bfnm{K.}\binits{K.}},
\bauthor{\bsnm{Toivonen},~\bfnm{H.~T.~T.}\binits{H.~T.~T.}} \AND
\bauthor{\bsnm{Korhola},~\bfnm{A.}\binits{A.}}
(\byear{2000}).
\btitle{A {B}ayesian multinomial {G}aussian response model for organism-based environmental reconstruction}.
\bjournal{J. Paleolimnol.}
\bvolume{24}
\bpages{243--250}.
\end{barticle}
%

\bptok{imsref}%
\endbibitem

%b46 ###
\bibitem[\protect\citeauthoryear{{Wiersma} and {Renssen}}{2006}]{WieRen06}
\begin{barticle}[author]
\bauthor{\bsnm{{Wiersma}},~\bfnm{A.~P.}\binits{A.~P.}} \AND
\bauthor{\bsnm{{Renssen}},~\bfnm{H.}\binits{H.}}
(\byear{2006}).
\btitle{{Model data comparison for the 8.2 ka BP event: Confirmation of a forcing mechanism by catastrophic drainage of Laurentide Lakes}}.
\bjournal{Quat. Sci. Rev.}
\bvolume{25}
\bpages{63--88}.\vadjust{\goodbreak}
\end{barticle}
%

\bptok{imsref}%
\endbibitem

%b47 ###
\bibitem[\protect\citeauthoryear{Woodward}{1987}]{Wood1987}
\begin{bbook}[author]
\bauthor{\bsnm{Woodward},~\bfnm{F.~I.}\binits{F.~I.}}
(\byear{1987}).
\btitle{Climate and Plant Distribution}.
%\bseries{Cambridge Studies in Ecology}.
\bpublisher{Cambridge Univ. Press},
\blocation{Cambridge}.
\end{bbook}
%

\bptok{imsref}%
\endbibitem
\end{thebibliography}
\end{document}